\documentclass[twocolumn]{aastex61}

\usepackage{upgreek}
\usepackage{amsmath}
\usepackage{hyperref}
\usepackage{threeparttable}

\received{July 1, 2016}
\revised{September 27, 2016}
\accepted{\today}
\submitjournal{ApJ}

\shorttitle{Exoplanet-induced radio emission}
\shortauthors{Turnpenney et al.}
\begin{document}

\title{Exoplanet-induced radio emission from M-dwarfs}

\correspondingauthor{Sam Turnpenney}
\email{st349@le.ac.uk}

\author[0000-0001-5010-7496]{Sam Turnpenney}
\affil{Department of Physics and Astronomy \\
University of Leicester \\
Leicester, LE1 7RH, UK}

\author[0000-0002-8004-6409]{Jonathan D. Nichols}
\affil{Department of Physics and Astronomy \\
University of Leicester \\
Leicester, LE1 7RH, UK}
%\collaboration{(AAS Journals Data Scientists collaboration)}

\author{Graham A. Wynn}
\affil{Department of Physics and Astronomy \\
University of Leicester \\
Leicester, LE1 7RH, UK}

\author{Matthew R. Burleigh}
\affil{Department of Physics and Astronomy \\
University of Leicester \\
Leicester, LE1 7RH, UK}
%\nocollaboration

%% Note that the \and command from previous versions of AASTeX is now
%% depreciated in this version as it is no longer necessary. AASTeX 
%% automatically takes care of all commas and "and"s between authors names.

%% AASTeX 6.1 has the new \collaboration and \nocollaboration commands to
%% provide the collaboration status of a group of authors. These commands 
%% can be used either before or after the list of corresponding authors. The
%% argument for \collaboration is the collaboration identifier. Authors are
%% encouraged to surround collaboration identifiers with ()s. The 
%% \nocollaboration command takes no argument and exists to indicate that
%% the nearby authors are not part of surrounding collaborations.

%% Mark off the abstract in the ``abstract'' environment. 
\begin{abstract}

We consider the magnetic interaction of exoplanets orbiting M-dwarfs, calculating the expected Poynting flux carried upstream along Alfv\'{e}n wings to the central star. A region of emission analogous to the Io footprint observed in Jupiter's aurora is produced, and we calculate the radio flux density generated near the surface of the star via the electron-cyclotron maser instability.  We apply the model to produce individual case studies for the TRAPPIST-1, Proxima Centauri, and the dwarf NGTS-1 systems. We predict steady-state flux densities of up to $\sim$ 10 $\upmu$Jy and sporadic bursts of emission of up to $\sim$ 1 mJy from each case study, suggesting these systems may be detectable with the Very Large Array (VLA) and the Giant Metrewave Radio Telescope (GMRT), and in future with the Square Kilometre Array (SKA).  Finally, we present a survey of 85 exoplanets orbiting M-dwarfs, identifying 11 such objects capable of generating radio emission above 10 $\upmu$Jy.

\end{abstract}

%% Keywords should appear after the \end{abstract} command. 
%% See the online documentation for the full list of available subject
%% keywords and the rules for their use.

%% From the front matter, we move on to the body of the paper.
%% Sections are demarcated by \section and \subsection, respectively.
%% Observe the use of the LaTeX \label
%% command after the \subsection to give a symbolic KEY to the
%% subsection for cross-referencing in a \ref command.
%% You can use LaTeX's \ref and \label commands to keep track of
%% cross-references to sections, equations, tables, and figures.
%% That way, if you change the order of any elements, LaTeX will
%% automatically renumber them.

%% We recommend that authors also use the natbib \citep
%% and \citet commands to identify citations.  The citations are
%% tied to the reference list via symbolic KEYs. The KEY corresponds
%% to the KEY in the \bibitem in the reference list below. 

\section{Introduction} \label{sec:intro}

Recently, seven terrestrial planets have been discovered orbiting the nearby ultracool dwarf star TRAPPIST-1 \citep{gillon2016, gillon2017}, and an Earth-sized planet, Proxima b, has been observed orbiting our nearest stellar neighbour, Proxima Centauri \citep{anglada2016}. These and other observations of planets orbiting M-dwarfs present an opportunity to study the magnetic interaction between extrasolar planets and their host stars. Proxima b and at least three of the planets at TRAPPIST-1 orbit in the nominal `habitable zone', defined as the region around a star in which liquid water may exist at the surface of a planet.  Many factors influence the potential habitability of a planet, and the existence of an intrinsic magnetic field may possibly play a vital role by protecting the atmospheres of exoplanets from erosive stellar radiation and winds \citep{khodachenko2007, lammer2009, seager2013, vidotto2013}. This is especially significant for M-dwarfs, since their habitable zones lie close to the star and planets would be subject to high XUV flux and stellar wind dynamic pressure \citep[e.g.][]{cohen2015, garraffo2017, wheatley2017}.% Exoplanet magnetic fields may also provide a means of revealing information about the internal structure of the planets, leading to characterisation of these bodies beyond bulk density.

Coherent, nonthermal radio emission attributed the electron-cyclotron maser instability (ECMI) is observed at magnetised planets in the solar system and is associated with auroral activity \citep{wu1979, zarka1992, ergun2000,treumann2006, imai2008}. Although auroral emission from exoplanets has yet to be detected \citep[e.g.][]{lazio2009, luger2017a}, the expected radio emission from them has been the subject of a number of studies. The `Radiometric Bode's Law' (RBL) is an empirical scaling relation  based on observations of radio emission from magnetised Solar System planets that is often extrapolated to estimate the radio power expected from exoplanets \citep[e.g.][]{lazio2004, zarka2007}. The RBL relates the output radio power from a planetary body to the Poynting or kinetic energy flux convected onto the obstacle, and has been used to demonstrate that hot Jupiters may be detectable with the next generation of radio telescopes \citep{farrell2004, lazio2004, griessmeier2007, zarka2007}. However, the RBL is empirical in nature and emission driven by ionospheric flow resulting from the reconnection of interplanetary and exoplanetary magnetic fields, analogous to the process generating Earth's auroral radio emission, has also been studied \citep{nichols2016}, and is predicted to be detectable with the next generation of radio telescopes. \citet{nichols2011} also examined the detectability of auroral radio emission from Jupiter-like exoplanets driven by currents generated by magnetosphere-ionosphere coupling, and identified 91 potential targets for radio detection within 25 pc of the Solar System \citep{nichols2012candidates}.  \citet{saur2013} considered the case of close-orbiting exoplanets in the sub-Alfv\'{e}nic regime, calculating the magnetic energy flux communicated away from the planet back towards the central star, and found that the total Poynting flux can reach values of $>$ 10$^{19}$ W in certain cases. Since habitable zone exoplanets around M-dwarfs orbit very close to the star, it is expected that the star-planet interactions there will be largely sub-Alfv\'{e}nic.  In light of the TRAPPIST-1 and Proxima Centauri b discoveries we compute radio emission powers due to M-dwarf-exoplanet interactions in the framework of \citet{saur2013} and estimate the radio emission generated at the star arising from the upstream transport of Poynting flux owing to the existence of a planetary obstacle to the stellar wind flow. Earth-sized exoplanets with plausible magnetic field strengths would produce radio emission directly from the planet that is below the ionospheric cutoff frequency of $\sim$10 MHz, therefore making it undetectable from Earth, regardless of the incident flux density \citep[e.g.][]{burkhart2017}. If, however, the relative velocity of a planetary body through a magnetised plasma is sub-Alfv\'{e}nic, then energy can be transported upstream of the flow along Alfv\'{e}n wings.  Jupiter's interaction with its Galilean satellites is a well known example of a sub-Alfv\'{e}nic interaction, producing powerful radio emissions and auroral footprints \citep{clarke1996, zarka1998, saur2004, jones2008, bonfond2009, wannawichian2010}. The interaction between Jupiter and its satellites has been discussed in depth by \citet{kivelson2004} and \citet{saur2004}. Similar interactions do not occur between solar system planets and the Sun, since the orbital distances are large enough that the solar wind is super-Alfv\'{e}nic in the planets' reference frame, prohibiting significant upstream-flow of energy. The interaction of exoplanets in close orbit with their host star, however, may be sub-Alfvenic, in which case emission analogous to the Io footprint and Io decametric (DAM) radio bursts may be produced near the surface of the star and modulated at the orbital period of the planet. Considerable attention has been given to the sub-Alfv\'{e}nic interaction of Jupiter with the Galilean satellites, serving as a useful benchmark for this work, due to the TRAPPIST-1 planetary system being comparable in scale to the Jovian satellite system.

In this paper we examine the detectability of radio emission resulting from sub-Alfv\'{e}nic star-planet interaction at M-dwarf exoplanet systems. We apply \citeauthor{saur2013}'s (\citeyear{saur2013}) analytic formulation for the magnetic energy communicated from an exoplanet to the central star via Alfv\'{e}n wings to three case studies, each providing a distinct motivation for examination.  Firstly, to TRAPPIST-1, a star hosting a system of multiple terrestrial exoplanets planets; secondly, to Proxima Centauri, the closest star to the Solar System, and host to a terrestrial exoplanet; and finally to NGTS-1, a more distant M-dwarf (224 pc) hosting a recently discovered hot Jupiter. In each case we estimate the radio flux generated by the ECMI from the surface of the star.  We also apply the same method to study a wider sample of exoplanet-hosting M-dwarfs, identifying those systems which may produce radio emission via the star-planet interaction that is detectable with current or next generation radio telescopes.

\section{Theoretical background}
\subsection{Sub-Alfv\'{e}nic interaction}

\begin{figure}
	\includegraphics[width=\columnwidth]{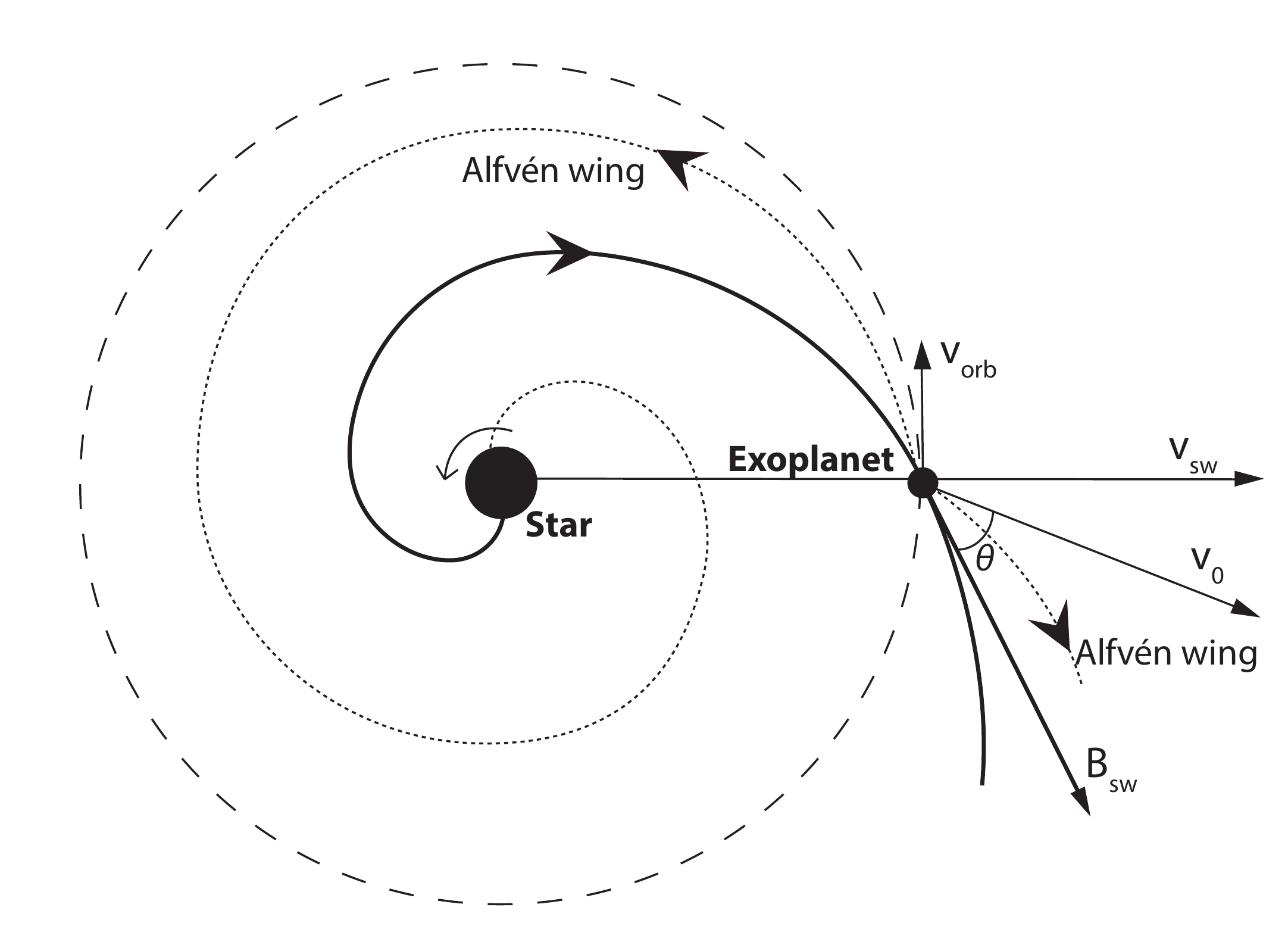}
    \caption{Schematic showing the geometry of the star-planet interaction. The two Alfv\'{e}n wings generated by the interaction are indicated by the dotted lines. The solid spiral emanating from the central star is the Parker spiral magnetic field. Highlighted are the stellar wind velocity $v_{\mathrm{sw}}$; the magnetic field of the stellar wind $B_{\mathrm{sw}}$; the orbital velocity $v_{\mathrm{orb}}$ of the exoplanet; the resultant velocity of the impinging stellar wind plasma $v_0$; and the angle $\theta$ between $v_0$ and $B_{\mathrm{sw}}$.}  
    \label{fig:schematic}
\end{figure}

A planetary body moving relative to an external magnetised plasma creates an obstacle to the plasma flow, thereby interacting with and modifying the surrounding environment \cite[see e.g.][and references therein]{saur2013}.  Of particular importance is the generation of Alfv\'{e}n waves in the wake of the flow, which are able to carry energy and momentum via the magnetic field.  If the velocity $v_0$ of the plasma relative to the planetary body is less than the Alfv\'{e}n speed $v_{\mathrm{A}}$, i.e. if 

\begin{equation} \label{cond1}
M_{\mathrm{A}} = \frac{v_0}{v_{\mathrm{A}}} < 1,
\end{equation}
where $M_{\mathrm{A}}$ is the Alfv\'{e}n Mach number, then the interaction generates two standing Alfv\'{e}n waves, making up Alfv\'{e}n wings \citep{neubauer1980}, which can propagate upstream of the flow, transporting energy and momentum in that direction (see Figure \ref{fig:schematic}). The condition for this case of the Alfv\'{e}n mode is satisfied in the solar system in the interaction of satellites with magnetised planets, and the sub-Alfv\'{e}nic interaction between Jupiter and its moons, particularly Io, has been widely studied \citep{prange1996, clarke1996, saur2004, jones2008, wannawichian2010}.  Alfv\'{e}n waves propagate along the magnetic field, and one of the wings, which are draped with respect to the field, is oriented back towards the central star provided the radial stellar wind velocity $v_{\mathrm{sw}}$ is less than the radial Alfv\'{e}n speed, i.e. if 

\begin{equation} \label{cond2}
v_{\mathrm{sw}} < \frac{B_r}{\left( \upmu_0 \rho_{\mathrm{sw}} \right)^{1/2}},
\end{equation}
where $B_r$ is the radial component of the interplanetary magnetic field (IMF), and $\rho_{\mathrm{sw}}$ is the mass density of the stellar wind. Equations (\ref{cond1}) and (\ref{cond2}) define the necessary conditions that must be met in order for the Poynting flux to communicate energy from the star-planet interaction back to central star.
 
\subsection{Cases of plasma flow-obstacle interaction}
 
There are at least four evident cases of plasma flow interaction with a planetary body.  These are (i) a magnetised planet with an atmosphere (e.g. Earth, Jupiter, Saturn), (ii) a magnetised planet with no atmosphere (e.g. Mercury), (iii) an unmagnetised planet possessing an atmosphere (e.g. Venus), and (iv) an unmagnetised, non-conducting body without an atmosphere (e.g. the Moon). In the first case of a magnetised planet with an atmosphere, currents due to the interaction with the stellar wind flow in the magnetosphere and Pedersen layer of the atmosphere, coupling the stellar wind to the planetary magnetosphere. In the case of an unmagnetised planet with an atmosphere the magnetopause is replaced by a magnetic pile-up boundary and a well defined plasma boundary.  The ionosphere replaces the magnetosphere acting as the boundary with which the IMF interacts in this case.  The case of a magnetised but airless planet still produces an interaction with the stellar wind, as is observed in the Sun-Mercury interaction, and in these cases the currents flow through the conducting region inside the planet (e.g. the iron core at Mercury). The final case of an unmagnetised, airless, non-conducting planetary body is expected to produce only a weak perturbation of the IMF in the wake of the body. In this paper we consider cases (i) and (iii), and the conclusions we reach apply to both magnetised and unmagnetised planets which posses atmospheres. We suggest the second  case for future work.

\subsection{Poynting flux within Alfv\'{e}n wings} 
 
The morphology envisaged for exoplanets in close orbit around M-dwarfs differs slightly from the Io-Jupiter interaction, in that the plasma flow is predominantly radial due to the outflow of stellar wind, rather than azimuthal as in the case of Io, and the expected geometry is shown in Figure \ref{fig:schematic}. The total Poynting flux carried within Alfv\'{e}n wings from an exoplanetary obstacle was calculated by \citet{saur2013} who also showed that for small Alfv\'{e}n Mach number the Poynting flux is approximated by

\begin{equation} \label{poyn}
S_{\mathrm{total}} = 2 \uppi R^2 \bar{\upalpha}^2 \frac{E_{\mathrm{sw}} B_{\perp}}{\upmu_0} M_{\mathrm{A}},
\end{equation}
where $R$ is the radius of the obstacle to the plasma flow, $E_{\mathrm{sw}}$ is the magnitude of the motional electric field (i.e. $\boldsymbol{E}_{\mathrm{sw}} = -\boldsymbol{v}_{\mathrm{sw}} \times \boldsymbol{B}_{\mathrm{sw}}$) seen in the rest frame of the planetary body, $\bar{\upalpha}$ denotes the strength of the sub-Alfv\'{e}nic interaction, and $B_{\perp}$ is the component of the stellar magnetic field perpendicular to the impinging plasma velocity at the orbital distance of the planet, given by 

\begin{equation}
B_{\perp} = B_{\mathrm{sw}} \sin \theta,
\end{equation}
where $\theta$ is the angle between the IMF $B_{\mathrm{sw}}$ and the incident stellar wind velocity, as shown in Figure \ref{fig:schematic}. The factor $\bar{\upalpha}$ represents the degree by which the motional electric field and plasma flow velocity are reduced by the interaction. For the cases considered in this paper, $\bar{\upalpha} = 1$ is a reasonable approximation, the justification for which is contained in Appendix A.

Note that we do not consider the exact nature of the flows within a magnetospheric or ionospheric obstacle.  The formulation developed by \citet{saur2013} employs the radius $R$ of the obstacle presented to the plasma flow in order to calculate the Poynting flux radiating away from the planet along interplanetary magnetic field lines due to the star-planet interaction. For an unmagnetised planet this is the radius of the ionosphere, which in this work we approximate as the planetary radius. For a magnetised planet we consider the radius of the magnetosphere, providing stellar wind and IMF conditions allow this to form.

The total Poynting flux, as expressed in equation (\ref{poyn}), and hence the spectral flux density, is a function of the magnetic and motional electric fields.  Values of these parameters are determined by the form of the stellar wind and magnetic field. The model of the stellar wind and magnetic field we employ is described in Appendix B. 

\subsection{Exoplanetary magnetic field strengths}

 To estimate unknown exoplanetary magnetic field strengths we employ the magnetic field scaling law proposed by \citet{sano1993}, whereby the planetary dipole moment $\mathcal{M}$ is estimated by

\begin{equation} \label{sano}
\mathcal{M} \, \, \propto\, \, \rho_c^{1/2} \Omega \ r_c^{7/2},
\end{equation}
where $\Omega$ is the rotational velocity of the planet, $\rho_c$ is the mass density in the dynamo region, for which we adopt the terrestrial value, and $r_c$ is the core radius of the planet, which can be determined from the mass of the planet by the empirical scaling law \citep{curtis1986}

\begin{equation}
r_c \, \, \propto \, \, M^{0.44}.
\end{equation}
The surface magnetic field strength of the planet can then be obtained by 

\begin{equation}  \label{magplnt}
B_p = \frac{\mathcal{M}}{R_p^3}.
\end{equation}
In the absence of any data on the rotation rates of exoplanets, we assume throughout this work that the planets are tidally locked, and thus that the rotational velocity is equal to the orbital velocity, justified by the extreme close orbits occupied by the majority of planets around M-dwarfs.  This assumption represents a lower limit on the planetary magnetic field, and a larger magnetic field would be calculated from this method if the planet has not yet become tidally locked.  Other scaling laws exist \citep[e.g.][]{busse1976, stevenson1983, mizutani1992} which provide similar planetary magnetic moment estimates as \citet{sano1993}.

\subsection{Radio power from near the stellar surface} \label{radio}

Providing the conditions of equations (\ref{cond1}) and (\ref{cond2}) are satisfied, then the total Poynting flux calculated from equation (\ref{poyn}) radiating away from the planet, is carried by one of the two Alfv\'{e}n wings back towards the central star (see Figure \ref{fig:schematic}). Shear Alfv\'{e}n waves carry magnetic field-aligned currents, which are associated with the ECMI, responsible for producing radio emission at Jupiter and Saturn.  We consider the radio emission generated near the surface of the star via the ECMI mechanism, at a frequency corresponding to the local gyrofrequency of the emission region \citep{treumann2006}. The total radio power emitted from one hemisphere of the star is given by

\begin{equation} \label{power}
P_r =\epsilon S_{\mathrm{total}},
\end{equation}
where $\epsilon$ represents the radio efficiency factor, i.e. the fraction of total Poynting flux converted to radio emission at the stellar surface.  To estimate $\epsilon$ we compare with Io-induced DAM radio emission at Jupiter.  \citet{saur2013} calculated that for Io the total Poynting flux from the magnetic interaction is $\sim$ 10$^{11}$ - $10^{12}$ W, while the observed Io DAM reaches 10$^9$ - 10$^{10}$ W \citep{clarke2004}, thus giving an overall radio efficiency of $\epsilon = 0.01$.  This represents a combination of the ECMI efficiency in addition to wave energy lost from the Alfv\'{e}n wing by reflection and conversion to other forms of energy, as is observed in sub-Alfv\'{e}nic interaction at Jupiter and Saturn \citep{wright1989, jacobsen2007, hess2011b}. This is also consistent with in situ measurements of the ECMI growth rate at Saturn \citep{lamy2011}, and for these reasons we employ this value of $\epsilon$ throughout our analysis.

Spectral flux density from the star can be obtained by

\begin{equation} \label{flux}
F_r = \frac{P_r}{\Omega \, s^2 \Delta \upnu},
\end{equation}
where $s$ is the distance from the star to the Earth, $\Delta \upnu$ is the bandwidth of the radio emission, and $\Omega$ is the solid angle into which the ECMI emission is beamed. No reliable estimates exist for the beaming angles of ECMI emission from low-mass stars. Hence, we assume throughout this work a reasonable value of $\Omega = 1.6$ sr, in conformity with observations of Jupiter's HOM and DAM emission \citep{zarka2004}, and also for the Io-related DAM emission \citep{ray2008}. The bandwidth is assumed to be equal to the gyrofrequency at the surface of the star in the equatorial region, an approximation consistent with solar system planet observations \citep{zarka1998}, and is hence given by

\begin{equation} \label{bandwidth}
\Delta \upnu = \frac{e B_{\star}}{2 \uppi m_e}, 
\end{equation}
where $B_{\star}$ is the average surface magnetic field strength of the star, and $e$ and $m_e$ are the charge and mass of the electron respectively. In this work we explore M-dwarf magnetic field strengths in the range 0.05 - 0.15 T, based on typical values measured at low-mass stars \citep{reiners2007, reiners2010, lynch2015}.  From equation (\ref{bandwidth}) this translates to an ECMI cutoff frequency range of $\sim$ 1.4  - 4.2 GHz.  However, we expect radio emission will be generated below and up to the cutoff frequency, and assume that the emission spectrum will be flat.  A number of radio telescopes are therefore be capable of testing the predictions proposed in this paper.   The Murchison Widefield Array (MWA) has achieved sensitivities of $\sim$ 10 mJy at 150 MHz \citep{tingay2013}, while the Giant Metrewave Radio Telescope (GMRT) is currently undergoing a major upgrade, which is expected to enable sensitivities of 190 $\upmu$Jy at 120 - 250 MHz, and 45 $\upmu$Jy at 1050 - 1450 MHz \citep{gupta2017}.  The Low Frequency Array (LOFAR) has achieved a flux density sensitivity of 10 mJy at 60 MHz \citep[e.g.][]{van2014}, while in the high-frequency regime the Very Large Array has obtained sensitivities of the order 10 $\upmu$Jy at 3 GHz \citep[e.g.][]{hallinan2017}.  Finally, the Square Kilometre Array (SKA) is a next generation radio telescope, due to begin observations in 2020, with an anticipated flux density sensitivity of $\sim$ 10 $\upmu$Jy \citep{zarka2015}. All of these telescopes will be considered when discussing the predictions of our model.

In the following section we apply the formulations described to three exoplanetary case studies, along with a wider survey of M-dwarf-exoplanet systems.

\section{Results}
\subsection{TRAPPIST-1} \label{trap}

We consider first the radio power expected from sub-Alfv\'{e}nic interactions between TRAPPIST-1 and its seven known exoplanets, orbiting between 0.011 and 0.063 AU with periods ranging between 1.51 and 18.77 days \citep{gillon2016, gillon2017, luger2017b}. The key stellar parameters are the rotation period, mass and radius of the star, 3.3 days, 0.08 $M_{\odot}$ and 0.117 $R_{\odot}$ respectively, in addition to the magnetic field strength and mass outflow rate of the star. The average surface magnetic field strength at TRAPPIST-1 was measured by \citet{reiners2010} as $B_{\star}$ = 0.06 T (where 1 Tesla = 10$^4$ Gauss). We employ a stellar mass outflow rate of $\dot{M_{\star}} = 1 \dot{M_{\odot}} = 3 \times 10^{-14} M_{\odot} \mathrm{yr}^{-1}$ \citep{garraffo2017}.

\begin{figure*}
	\includegraphics[width=0.8\textwidth]{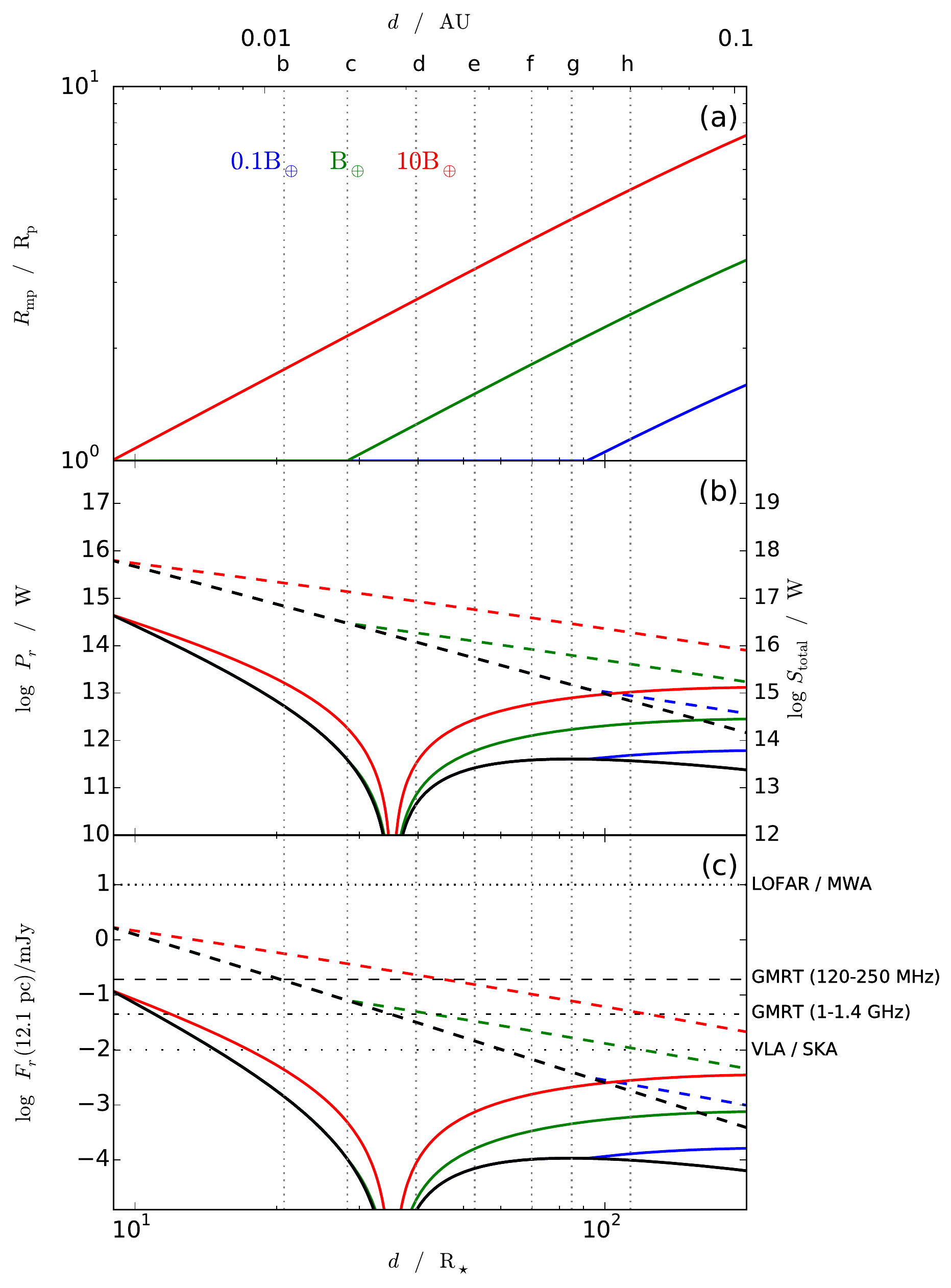}
    \caption{Plots of derived parameters for the TRAPPIST-1 exoplanetary system versus orbital distance $d$ in units of stellar radii $R_{\star}$ and AU for intrinsic planetary magnetic field strengths of 0.1 $B_{\oplus}$ (blue lines), 1 $B_{\oplus}$ (green lines) and 10 $B_{\oplus}$ (red lines). Solar values of the XUV luminosity $L_{\mathrm{XUV}}$ and stellar wind mass outflow rate $\dot{M}$ were used, along with a planetary radius of 1 $R_{\oplus}$ for all seven planets. Panel (a) shows the magnetopause standoff distance $R_{\mathrm{mp}}$ in units of planetary radii.  Panel (b) shows the radio power $P_{\mathrm{r}}$ from the star-planet interaction (left axis) and the corresponding total Poynting flux (right axis) for the idealised Parker spiral (solid lines), and the maximum power of radio bursts (dashed lines), with the black lines representing the case of an unmagnetised planet. Finally, panel (c) shows the radio flux density $F_{\mathrm{r}}$ from the surface of the star in units of mJy, assuming a distance from Earth of 12.1 pc. The horizontal lines indicate sensitivity thresholds of the respective radio telescopes.}  
    \label{fig:trap_derived}
\end{figure*}

We now derive parameters characterising the interaction of the exoplanets with the stellar wind conditions.  For the purpose of plotting, we have assumed that each of the exoplanets around TRAPPIST-1 has a radius of 1 R$_{\oplus}$, while Table. \ref{tab:trappist} shows the derived parameters for the individual planets using their measured masses and radii.  Firstly, in Figure \ref{fig:trap_derived}(a) we show the size of the magnetopause standoff distance, calculated using equation (\ref{rmp}), for a range of three plausible planetary magnetic field strengths bracketing the terrestrial value.  The results show that for an intrinsic planetary field strength of 10 B$_{\oplus}$, small magnetospheres ranging in magnetopause standoff distance of $\sim$1.1 - 1.4 R$_{\mathrm{P}}$ form at all seven planets. For a terrestrial field strength only the outer five planets form magnetospheres, while a planetary field strength of 0.1 B$_{\oplus}$ precludes the formation of a magnetosphere at six of the seven exoplanets.

Turning now to the ECMI-induced radio power from the surface of the star calculated using equations (\ref{poyn}) and (\ref{power}), where the additional black lines in Figure \ref{fig:trap_derived}(b) show the power resulting from unmagnetised exoplanets, i.e. where the radius $R$ of the obstacle is simply the radius of the planet. The solid lines represent the lower limit of the power in the idealised case of a perfect Parker spiral, i.e. where there is no $B_z$ component of the magnetic field, only radial $B_r$ and azimuthal $B_{\phi}$ components.  However, this is unrealistic, since a number of factors, such as a magnetic field tilted with respect to the planetary orbit, the occurrence of transient stellar activity such as coronal mass ejections, or a tilting of the current sheet, would result in deviation from the idealised Parker spiral. Hence, the `notch' in the vicinity $35 R_{\star}$, which occurs when the incident plasma velocity is parallel to the IMF, would not actually drop to zero, and even if the above mentioned factors were not present, we would still expect interaction due to viscous forces which would slow the stellar wind and produce Alfv\'{e}n wings. Therefore, we have also shown with the dashed coloured lines in Figure \ref{fig:trap_derived}(b) the upper limit of the radio power, which is calculated using the total IMF strength $B_{\mathrm{sw}}$ in equation (\ref{poyn}) rather than the component perpendicular to the incident stellar wind velocity.  The magnetic field actually experienced will lie somewhere between the two.  The dashed lines therefore represent the upper limit for bursts of radio power resulting from intervals of high magnetic field strength above the steady state value represented by the solid lines.  We observe that the steady state radio power (solid lines) induced by sub-Alfv\'{e}nic interaction peaks at the orbits of the innermost and outmost planets around TRAPPIST-1, at $\sim$ 10$^{13}$ W, although sporadic bursts may reach radio powers of up to two orders of magnitude greater.  By comparison, the radio power from the Io DAM emission at Jupiter is $\sim$ 10$^9$ - 10$^{10}$ W \citep{prange1996, clarke1996, gerard2006}.  Figure \ref{fig:trap_derived}(c) shows the spectral flux density from TRAPPIST-1 given by equation (\ref{flux}).  The horizontal lines running across the plot indicate the sensitivities of the radio telescopes, as discussed in Section \ref{radio}. With steady state flux densities of up to $\sim$ 10 $\upmu$Jy, and bursts producing flux densities of almost 1 mJy, our results show that we do not expect steady state emission to be presently detectable, assuming plausible planetary magnetic fields.  However, planets with field strengths greater than 1 $\mathrm{B_{\oplus}}$ may induce sporadic radio bursts which are detectable with the VLA and in the future with the SKA.  The GMRT may also be capable of detecting sporadic bursts if the planetary field strength is $\sim$ 10  $\mathrm{B_{\oplus}}$. We note, however, that the planetary magnetic field strengths shown in Table \ref{tab:trappist}, which were calculated using \citeauthor{sano1993}'s (\citeyear{sano1993}) scaling law, preclude the formation of magnetospheres at any of the seven planets.  Hence, in this case the flux density would be that indicated by the black line in Figure \ref{fig:trap_derived}(c), implying that radio bursts from the four innermost planets only would be detectable with the VLA or SKA. \citet{garraffo2017} estimate magnetopause standoff distances of up to $\sim$ 2.7 $R_p$ at the outermost exoplanets, since they adopt Earth-like planetary magnetic field strengths. In contrast, \citeauthor{sano1993}'s (\citeyear{sano1993}) law yields the smaller planetary magnetic field strength found in this work due to the smaller masses and longer rotation periods of the planets compared with Earth.

We note that the coronal temperature of 2 MK employed in our model is an estimate of a weakly constrained quantity which we have adopted following the work of \citet{vidotto2011, vidotto2013}. However, coronal temperatures at M-dwarfs may reach 10 MK \citep{schmitt1990, giampapa1996}, therefore we also examined the effects of a 10 MK corona, corresponding to a stellar wind sound speed of 379 km s$^{-1}$.  A temperature of 10 MK is also the value which corresponds to stellar wind sound speeds consistent with those of \citet{garraffo2017}.  We find that the resulting radio powers and flux densities are reduced by less than one order of magnitude for this higher coronal temperature and sound speed. 

\begin{figure*}
	\includegraphics[width=\textwidth]{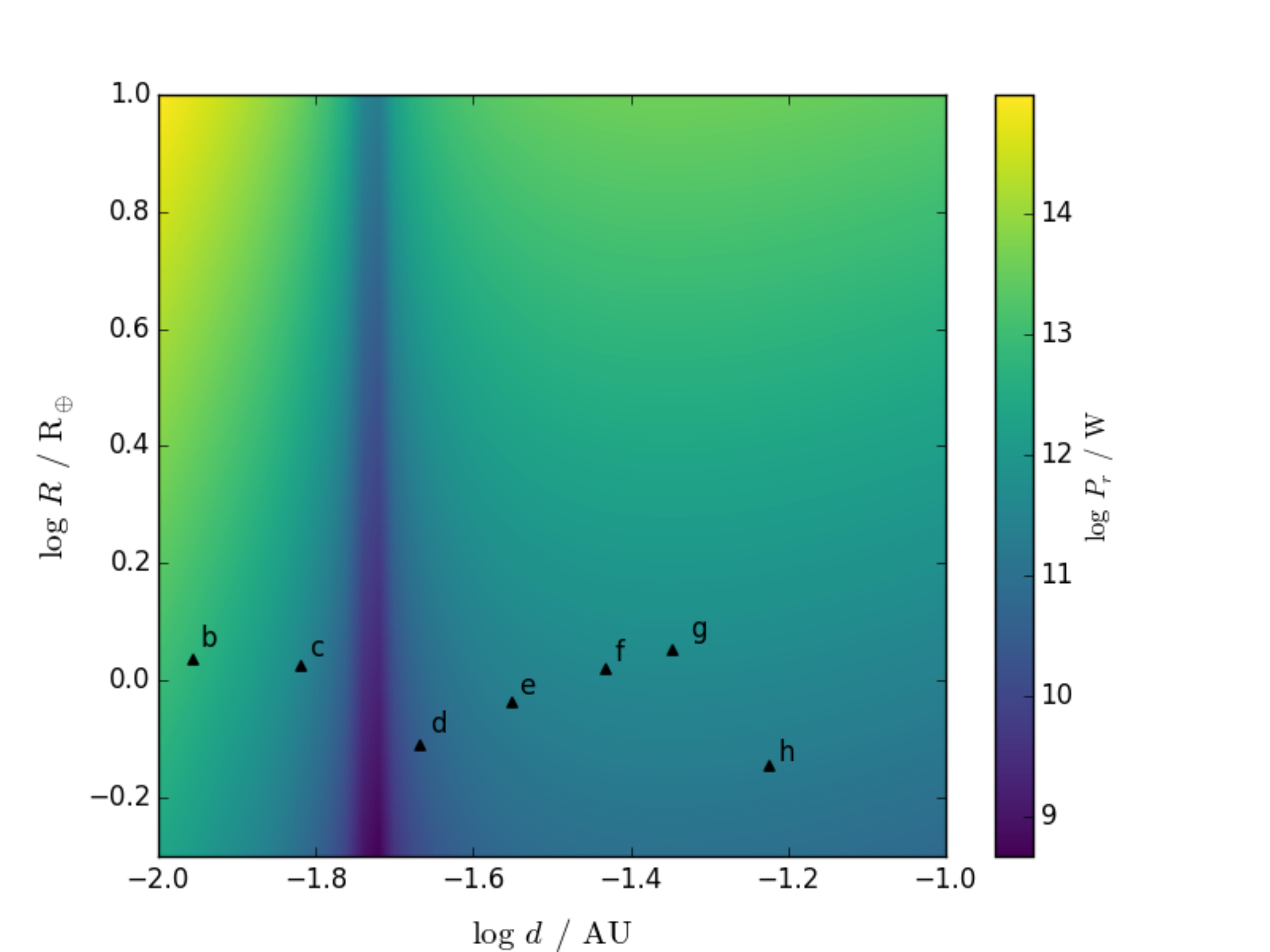}
    \caption{A color plot for TRAPPIST-1 of radio power $P_r$ versus distance from the central star $d$ in units of AU, and the radius of the obstacle to the plasma flow $R$ in units of Earth radii. Markers indicate the positions of the exoplanets, where the measured radii, as in Table \ref{tab:trappist}, have been used.}  
    \label{fig:cont_trap}
\end{figure*}

In Figure \ref{fig:cont_trap} we show a map of radio power $P_r$ versus orbital distance $d$ and radius of the obstacle $R$.  Apparent is the general trend that radio power increases with obstacle radius, and decreases with orbital distance.  This plot has been produced for an idealised Parker spiral IMF, and thus the `notch' evident in Figure \ref{fig:trap_derived} is also a feature here at $\sim$ 0.02 AU.  Such color plots may be used in conjunction with future radio detections to infer the radii of detected obstacles, assuming the planetary orbital radius and relevant stellar parameters are known.  Significant persistent deviations from the points plotted (corresponding to known radii) will indicate the existence and strength of planetary magnetic fields.  Multiple planetary systems such as TRAPPIST-1 will provide the tightest constraints, and similar plots can in principle be produced for many stars.

\begin{deluxetable*}{cccccccccccc}
%\tablenum{2}
\tablecaption{Properties of the planets orbiting TRAPPIST-1\label{tab:trappist}}
\tablewidth{0pt}
\tablehead{
\colhead{Planet} & \colhead{$d$\textsuperscript{a, b}} & \colhead{Orbital period\textsuperscript{a, b} } & \colhead{$n_{\mathrm{sw}}$\textsuperscript{c} } &
\colhead{$v_{\mathrm{sw}}$\textsuperscript{d} } & \colhead{$B_{\mathrm{sw}}$\textsuperscript{e} } & \colhead{$M_p$\textsuperscript{f} }&
\colhead{$R_p$\textsuperscript{a} } & \colhead{$B_p$\textsuperscript{g} } & \colhead{$R_{\mathrm{mp}}$  / $R_p$\textsuperscript{h} } &
\colhead{$P_r$\textsuperscript{i} } & \colhead{$F_r$\textsuperscript{j} } \\
\nocolhead{Number} & \colhead{(AU)} & \colhead{(days)} & \colhead{(cm$^{-3}$)} &
\colhead{(km$^{-1}$)} & \colhead{(nT)} & \colhead{($M_{\oplus}$)} &
\colhead{($R_{\oplus}$)} & \colhead{($B_{\oplus}$)} & \nocolhead{} & \colhead{(W)} & \colhead{($\upmu$Jy)}
}
%\decimalcolnumbers
\startdata
b  & 0.0111  & 1.511 & 38519 &492 & 137169 & 0.79 & 1.086 & 0.353 &1& 5.92 $\times$ 10$^{12}$ & 1.58 \\
c  & 0.0152  & 2.422 & 19131 &528 & 73276 & 1.63 & 1.056 & 0.731 &1& 3.71 $\times$ 10$^{11}$ & 0.099\\
d  & 0.0215  & 4.050 & 8923  &566 & 36746 & 0.33 & 0.772 & 0.096 &1& 3.56 $\times$ 10$^{10}$ & 0.00949\\
e  & 0.0282  & 6.100 & 4942  &594 & 21451 & 0.24 & 0.918 & 0.023 &1& 2.26 $\times$ 10$^{11}$ & 0.0603\\
f  & 0.0371  & 9.206 & 2730  &621 & 12479 & 0.36 & 1.045 & 0.019 &1& 4.22 $\times$ 10$^{11}$ & 0.113\\
g  & 0.0451  & 12.353& 1793  &640 & 8506 & 0.566& 1.127 & 0.023 &1& 5.14 $\times$ 10$^{11}$ & 0.137\\
h  & 0.0596  & 18.766& 987   &666 & 4947 & 0.086& 0.715 & 0.0033&1& 1.91 $\times$ 10$^{11}$ & 0.0510\\
\enddata
\tablecomments{\textsuperscript{a} From \citet{gillon2017}; \textsuperscript{b} From \citet{luger2017b}; \textsuperscript{c} From equation (\ref{number}); \textsuperscript{d} From equation (\ref{sw}); \textsuperscript{e} From equation (\ref{mag}); \textsuperscript{f} From \citet{wang2017}; \textsuperscript{g} From equation (\ref{magplnt}); \textsuperscript{h} From equation (\ref{rmp}) but equal to unity if $R_{\mathrm{mp}} < R_{\mathrm{p}}$; \textsuperscript{i} From equation (\ref{power}); \textsuperscript{j} From equation (\ref{flux}). \\}
\end{deluxetable*}

\subsection{Proxima Centauri}

As a second case study we also examine the radio emission from Proxima Centauri due to sub-Alfv\'{e}nic interaction with its exoplanet, Proxima b, which orbits at a distance of 0.04 AU. The planet has a minimum measured mass of 1.3 $\mathrm{M_{\Earth}}$, implying a radius of 1.1 $\mathrm{R_{\Earth}}$ assuming a density equal to that of Earth.  A number of crucial factors differentiate the results for Proxima Centauri compared with TRAPPIST-1, namely: the much longer rotation period of Proxima Centauri, 82.6 days; the larger mass and radius of Proxima Centauri, 0.122 $M_{\odot}$ and 0.154 $R_{\odot}$ respectively; and the distance of Proxima Centauri to the Earth, just 1.3 pc compared with 12.1 pc for TRAPPIST-1 \citep{anglada2016}. The measured average surface magnetic field strength of Proxima Centauri is 0.06 T \citep{reiners2008}, and the mass outflow rate from is estimated to be similar to the solar value, therefore we again use $\dot{M_{\star}} = 3 \times 10^{-14} M_{\odot} \mathrm{yr}^{-1}$ \citep{garraffo2016}.

\begin{figure*}
	\includegraphics[width=0.8\textwidth]{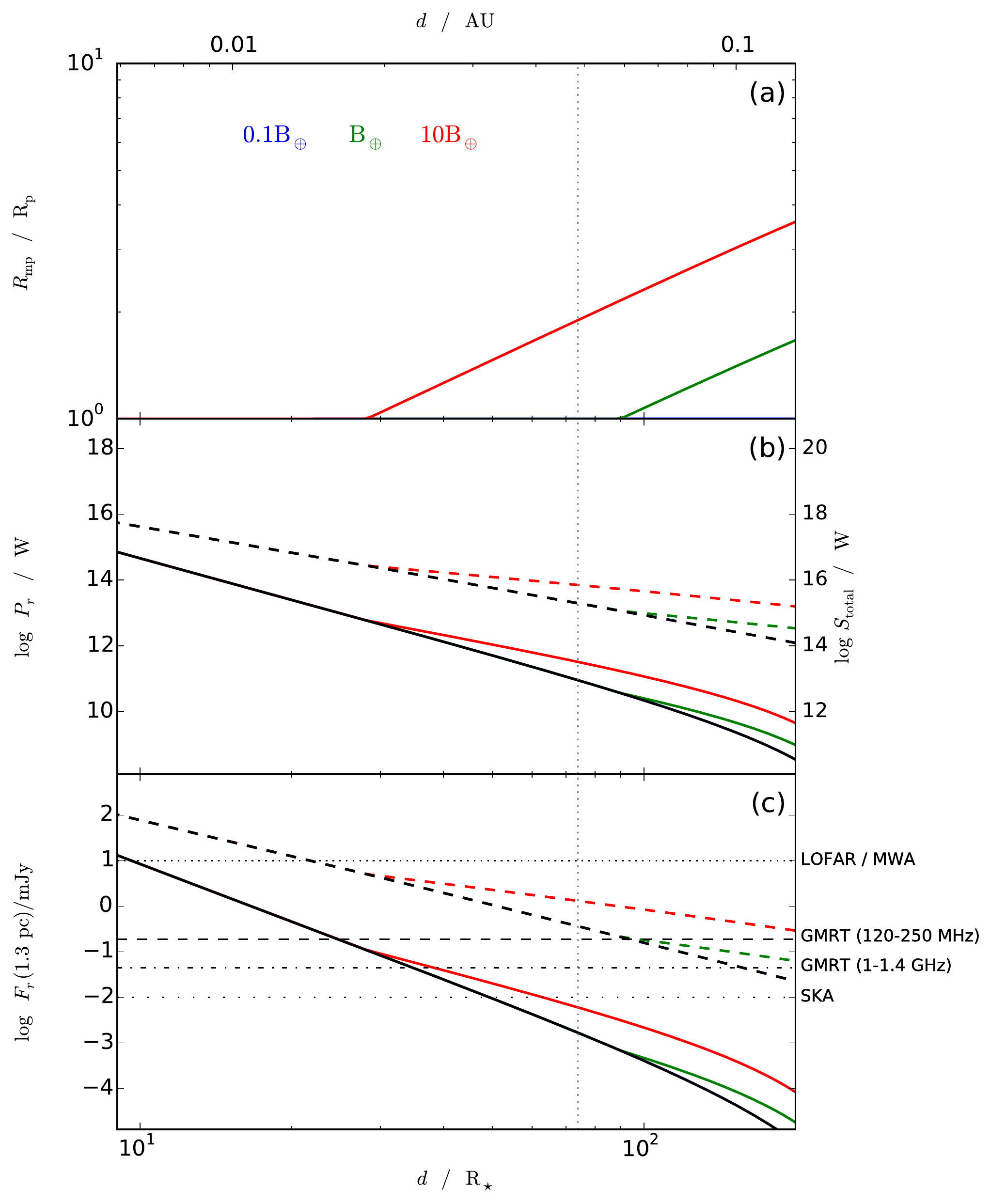}
    \caption{As for Figure \ref{fig:trap_derived} but for Proxima Centauri b. Again solar values of $L_{\mathrm{XUV}}$ and $\dot{M}$ were used, along with a planetary radius of 1.1 $R_{\oplus}$}
    \label{fig:proxima_derived}
\end{figure*}

Figure \ref{fig:proxima_derived} shows the derived parameters taking a planetary magnetic field strength of $0.1 \; \mathrm{B_{\oplus}}$, close to the exact value of $0.0952 \; \mathrm{B_{\oplus}}$ calculated from equations (\ref{sano}) - (\ref{magplnt}), and bracketing this with values an order of magnitude either side. We see from Figure \ref{fig:proxima_derived}(a) that a magnetic field strength greater than the value of $0.1 \mathrm{B_{\oplus}}$ given by \citeauthor{sano1993}'s (\citeyear{sano1993}) scaling law is required in order to form a magnetosphere at the planet. The radio power from Proxima Centauri shown in Figure \ref{fig:proxima_derived}(b) has a value of $\sim 10^{12}$ W assuming the terrestrial value for the field strength, and $\sim 10^{11}$ W if the planetary field strength is low enough to prohibit the formation of a magnetosphere, while radio bursts of up to $\sim 10^{14}$ W are plausible.  Despite lower radio power, the flux density from Proxima Centauri (Figure \ref{fig:proxima_derived}(c)) is slightly larger than that from TRAPPIST-1 owing to the system's much closer proximity to Earth, yielding a value of $\sim \; 10 \;\upmu$Jy for a terrestrial magnetic field strength. Radio bursts producing flux densities of up to $\sim$ 1 mJy are also possible. Hence, coherent radio emission from sporadic bursts from Proxima Centauri may be detectable by the GMRT and SKA. We note that with a declination of approximately -60$^{\circ}$, Proxima Centauri is not visible at the latitude of the VLA.  \citet{bell2016} observed Proxima Centauri with the Murchison Widefield Array, resulting in a non-detection.  They place an upper limit on radio emission from the system at 42.3 mJy/beam (3-sigma) at 200 MHz, a limit at least two orders of magnitude greater than the flux density predicted by our model.

\begin{figure*}
	\includegraphics[width=\textwidth]{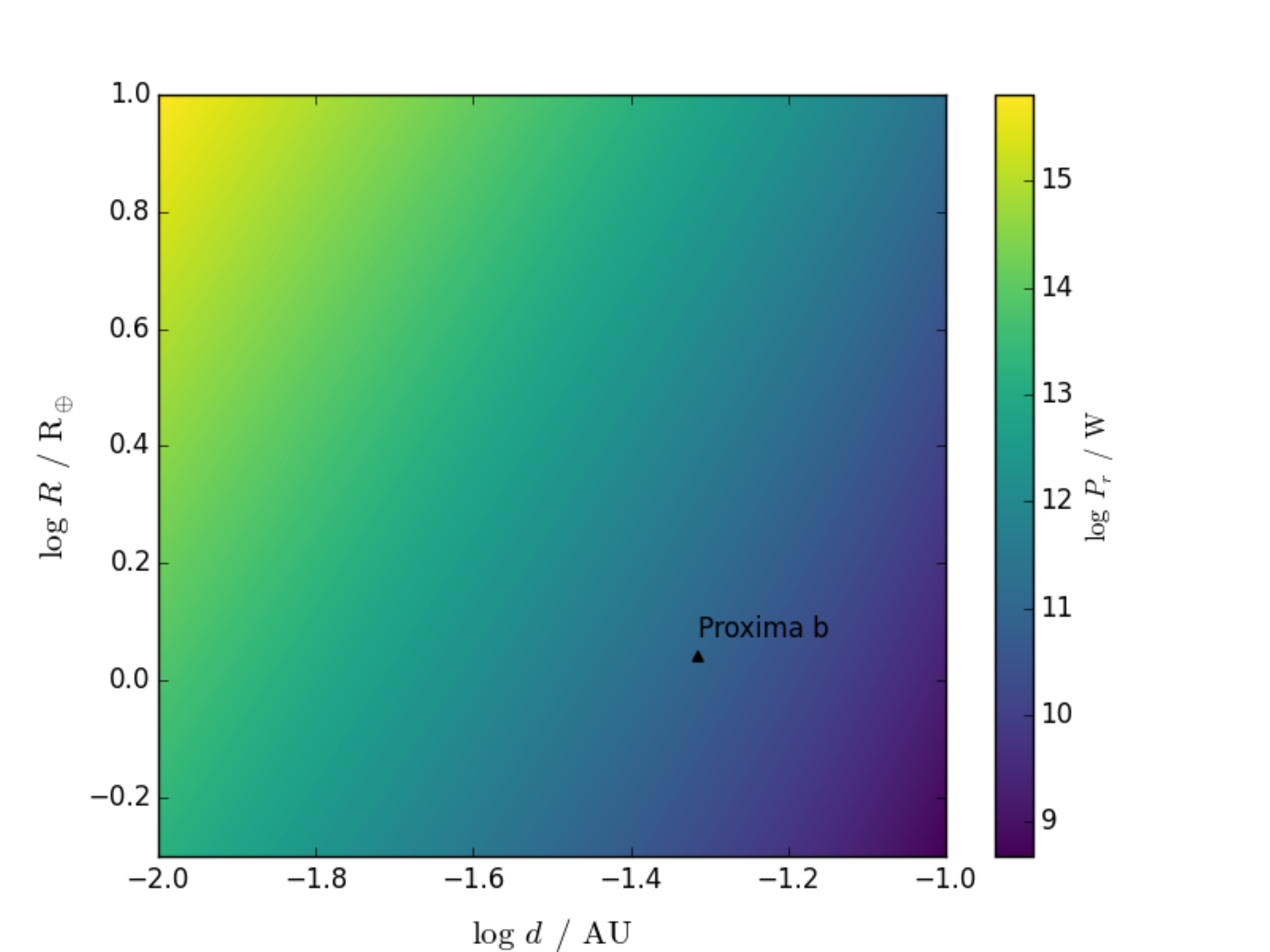}
    \caption{As for Figure \ref{fig:cont_trap}, but with Proxima Centauri stellar parameter values.} 
    \label{fig:cont_proxima}
\end{figure*}

Figure \ref{fig:cont_proxima} shows a color plot of the radio power $P_r$ from the star-planet interaction.  A noticeable difference from the equivalent plot for TRAPPIST-1 (Figure \ref{fig:cont_trap}) is the lack of the `notch' in the results.  This is due to the much slower rotation of Proxima Centauri, meaning that the incident plasma velocity does not become parallel to the IMF until a greater orbital distance, beyond the range of the plot.  As with TRAPPIST-1 this power map may be used to determine the obstacle radius and thus the nature of any planetary magnetic field at Proxima-b, by comparison with any future radio emission detected.

\subsection{NGTS-1}

The previous two case studies, Proxima Centauri and TRAPPIST-1, are systems containing approximately Earth-sized planets.  For a final case study we examine a system comprising a hot Jupiter in orbit around an M-dwarf, situated 224 pc from Earth.  NGTS-1b is a planet of mass  0.812 $\mathrm{M_{Jup}}$ and radius 1.33 $\mathrm{R_{Jup}}$, orbiting an M0.5 dwarf of mass 0.617 $\mathrm{M}_{\odot}$ and radius 0.573 $\mathrm{R}_{\odot}$ in a P = 2.647 d orbit \citep{bayliss2017}.   The magnetic field strength of NGTS-1 is unknown, and therefore we have assumed a value of 0.06 T, consistent with the two previous case studies.  A stellar rotation period of 43 days is estimated from photometric analysis of spot coverage \citep{bayliss2017}. In common with our studies of TRAPPIST-1 and Proxima Centauri, we have assumed $L_{\mathrm{XUV}} = L_{\mathrm{XUV \odot}}$ at NGTS-1.

\begin{figure*}
	\includegraphics[width=0.8\textwidth]{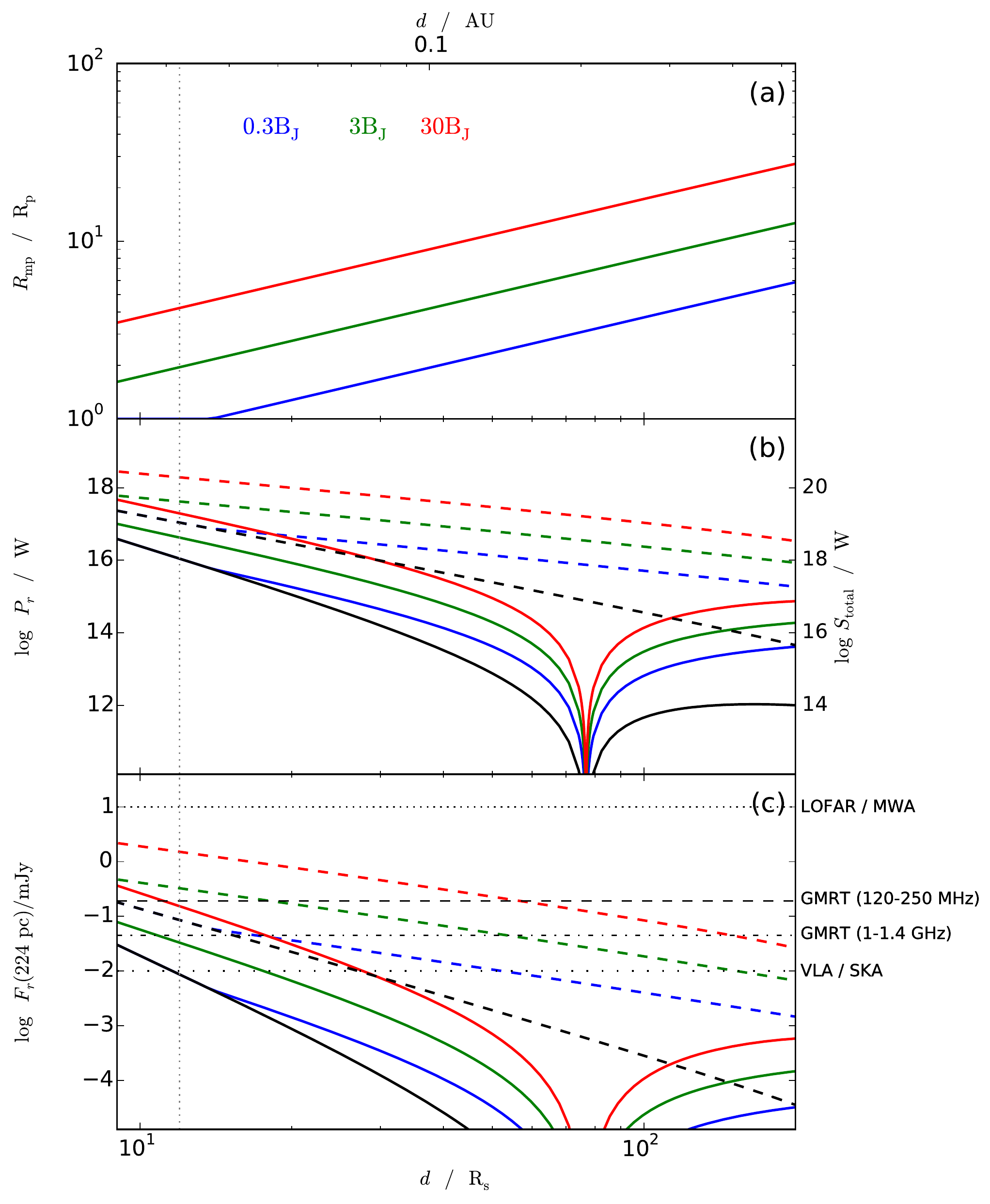}
    \caption{As for Figures \ref{fig:trap_derived} and \ref{fig:proxima_derived} but for NGTS-1 and with the following difference: Planetary magnetic field strengths of 0.3 $B_J$ (blue lines), 3 $B_J$ (green lines) and 30 $B_J$ (red lines) are plotted.}
    \label{fig:nltt_derived}
\end{figure*}

 \begin{figure*}
	\includegraphics[width=\textwidth]{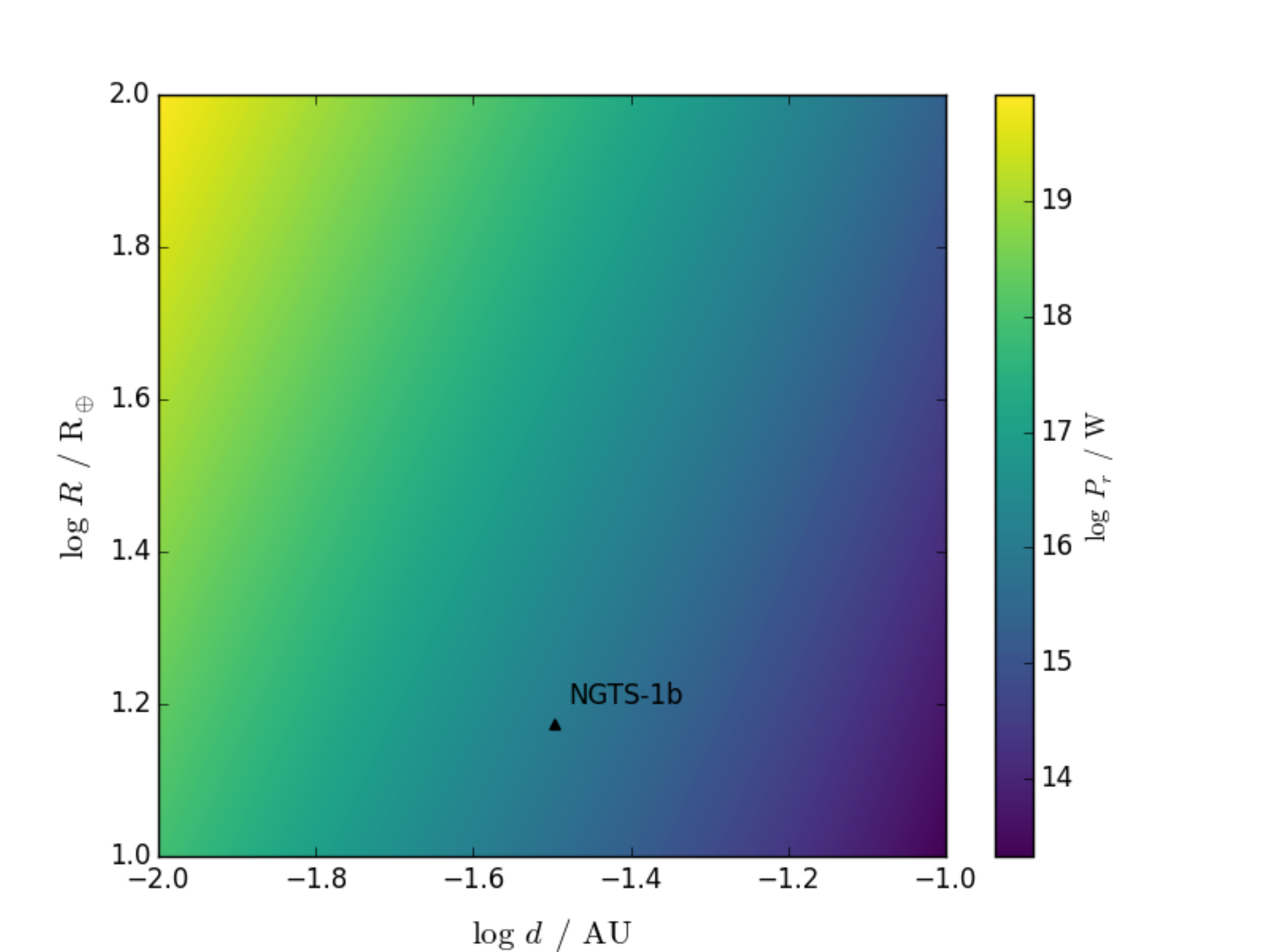}
    \caption{As for Figures \ref{fig:cont_trap} and \ref{fig:cont_proxima}, but with NGTS-1 stellar parameter values. Here we now consider an obstacle radius $R$ range of 10 - 100 $\mathrm{R_{\oplus}}$ to reflect the fact that the exoplanet in this system is a hot Jupiter of radius 1.33 $\mathrm{R_{Jup}}$.} 
    \label{fig:cont_ngts}
\end{figure*}

The derived planetary parameters are shown in Figure \ref{fig:nltt_derived}. For this particular case study of a hot Jupiter, we use a scaling law appropriate for giant planets in the form given by \citet{reiners2009} to estimate the strength of the magnetic field of NGTS-1b:

\begin{equation} \label{dyn}
B_{\mathrm{dyn}} = 4.8 \times \left( \frac{ML^2}{R^7} \right)^{1/6}   [\mathrm{kG}],
\end{equation}
where $B_{\mathrm{dyn}}$ is the mean strength of the magnetic field at the surface of the dynamo, and $M$, $L$, and $R$ are the mass, luminosity, and radius of the planet respectively, all normalized with solar values.  The equatorial dipole field strength is related to the field strength at the surface of the dynamo by \citep{reiners&christensen2010}

\begin{equation} \label{dip}
B_p = \frac{B_{\mathrm{dyn}}}{2 \sqrt{2}} \left(1 - \frac{0.17}{M/M_J} \right)^3.
\end{equation}
We use in equation (\ref{dyn}) a typical luminosity value of $L = 10^{-7} L_{\odot}$  taken from the evolutions tracks calculated by \citet{burrows1997}. For NGTS-1b, equations (\ref{dyn}) and (\ref{dip}) give an equatorial field strength of $B \simeq 1.5$ mT, i.e. approximately three times greater than the surface field strength of Jupiter, and we also examine field strengths an order of magnitude either side of this value. Figure \ref{fig:nltt_derived}(a) shows that a magnetopause may form with a stand-off distance $\leq 4 \; \mathrm{R_{p}}$ at the orbital radius.  Figure \ref{fig:nltt_derived}(b) shows that the radio power carried in the Alfv\'{e}n wing is $\sim 10^{16} - 10^{18}$ W, which translates to a flux density from the surface of the M-dwarf (Figure \ref{fig:nltt_derived}(c)) of $\sim$ 0.01 mJy - 1 mJy. These results suggest that a planetary field strength at the upper end of the range considered may produce emission that is detectable by the VLA, GMRT, and SKA. A color plot of radio power is shown in Figure \ref{fig:cont_ngts}, which may again enable determination of planetary magnetic field strength from future radio observations, via the method described in Section \ref{trap}. We note that planets such as NGTS-1 which potentially possess magnetic fields comparable to Jupiter in strength are important, since they may in principle be also detectable from the ground via direct planetary radio emissions. Comparison of stellar and planetary emissions will yield vital information regarding the nature of the interaction over and above the scope of the present model.

\subsection{M-dwarf exoplanets}

We now present analysis of a survey of M-dwarf-exoplanet systems using data listed in the \url{https://exoplanetarchive.ipac.caltech.edu} and \url{http://exoplanet.eu} catalogues. Selecting only those exoplanets orbiting within 1 AU of the host star, we identify a total of 85 exoplanets for inclusion in our study, and calculate the expected radio flux for each via sub-Alfv{\'e}nic magnetic interaction.  As discussed above, the properties required to calculate the flux density are: the incident stellar wind velocity $v_0$; stellar magnetic field components $B_r$ and $B_{\phi}$; the density $\rho_{\mathrm{sw}}$ of the stellar wind; and the radius of the planetary obstacle $R$. None of these quantities are directly measurable at present. Therefore, just as with our case study results, we estimate the required quantities as described in  Appendix B. 

Unlike the magnetic field strengths of TRAPPIST-1 and Proxima Centauri which have reasonably well constrained estimates, the magnetic field strengths of the other M-dwarfs in our study are, for the most part, unknown.  Typical low-mass star magnetic field strengths are commonly of the order 0.1 T \citep[e.g.][]{reiners2007, lynch2015}. Therefore, in the absence of any data, we have applied our model for three magnetic field strengths of $B_{\star} = 0.05,\, 0.1,\, 0.15$ T for all stars in the survey.  To estimate the effective radius of the obstacle, knowledge of the intrinsic planetary magnetic field is required. In the case studies we examined a range of three separate values for the magnetic field strength.  However, in this survey we desire a single flux density value for each system, and we therefore again employ the scaling law of \citet{sano1993} to arrive at a magnetic field strength value given by equations (\ref{sano}) - (\ref{magplnt}). As with the case studies of TRAPPIST-1 and Proxima b, we assume that all exoplanets in this study are tidally locked; an assumption likely to be valid in most cases due to the close orbits and ages of the majority of systems.

In the cases where either database is incomplete, we estimate the values of any missing properties as follows: If the radius of the star $R_{\star}$ is missing, we estimate the value based on typical values for the spectral type of the star, as given in Appendix G of \citet{carroll2007}.  If either the radius $R_{\mathrm{p}}$ or mass $M_{\mathrm{p}}$ of the exoplanet is unknown, we assume a terrestrial density to estimate the missing quantity. In case the orbital period $P_{\mathrm{orb}}$ is unmeasured, we use Kepler's third law to calculate the missing value. Since data on rotation rates of M-dwarfs is sparse, we have assumed a period of 19 days for all dwarfs based on results from a survey of M-dwarf rotation periods by \citet{mcquillan2013}.  Additionally, we have assumed throughout a stellar mass loss rate of $\dot{M_{\star}} = \dot{M_{\odot}}$, based on the mass loss rates of a similar magnitude used in our earlier case studies.

We calculate the radio power and flux density for the selected targets using the same method as in the earlier case studies. Of the 85 M-dwarf-orbiting exoplanets, 11 are estimated to generate radio flux above 10 $\upmu$Jy, and the results for these targets are shown in Table \ref{tab:mdwarfs}. Only one of the exoplanets generating potentially detectable emission is estimated to possess a magnetosphere, and thus in the majority of cases the effective radius of the planetary obstacle to the stellar wind is simply the planetary radius. We note that the results show that flux density is independent of stellar field strength $B_{\star}$ for cases where no magnetosphere forms, i.e. where $R_{\mathrm{mp}} / R_p = 1$.  Closer inspection of equation (\ref{poyn}) reveals that although $S_{\mathrm{total}} \propto B_{\star}$, this is cancelled by a 1/$B_{\star}$ term in the flux density calculation arising from our assumption that the bandwidth of the emission is the cyclotron frequency in the generation region at the surface of the star. %in the ECMI emission region, which we take to be at the stellar surface.

\begin{longrotatetable}
\begin{deluxetable*}{ccccccccccc}
\tablenum{2}
\tablecaption{Flux density estimates for exoplanets orbiting M-dwarfs\label{tab:mdwarfs}}
\tablewidth{700pt}
\tablehead{
\colhead{Exoplanet} & \colhead{Spectral} & \colhead{$d$} & \colhead{$M_{\star}$} &
\colhead{$R_{\star}$} & \colhead{$M_{\mathrm{p}}$} & \colhead{$B_{\mathrm{p}}$} &
\colhead{$s$} & \colhead{$R_{\mathrm{mp}} / R_{\mathrm{p}}$} & \colhead{$P_r [B_{\star}=0.05, 0.1, 0.15 \mathrm{T}]$} & \colhead{$F_r$} \\
\nocolhead{Number} & \colhead{type} & \colhead{(AU)} & \colhead{($M_{\odot}$)} &
\colhead{($R_{\odot}$)} & \colhead{($M_{\mathrm{Jup}}$)} & \colhead{(nT)} &
\colhead{(pc)} & \nocolhead{} & \colhead{(W)} & \colhead{($\upmu$Jy)}
}
%\decimalcolnumbers
\startdata
GJ 876 d          & M4     & 0.021  &  0.33  & 0.35 & 0.021 &4.38 $\times 10^4$  & 4.7 &1  & 1.44 $\times 10^{14}$, 2.89 $\times 10^{14}$, 4.33 $\times 10^{14}$& 306.6\\
GJ 436 b          & M2.5 V & 0.029  &  0.47  & 0.46 & 0.07  &2.00 $\times 10^4$  & 10.23&1 & 4.02 $\times 10^{14}$, 8.04 $\times 10^{14}$, 1.21 $\times 10^{15}$& 180.1\\
		GJ 674 b          & M2.5   & 0.039  &  0.35  & 0.48 & 0.035 &2.38 $\times 10^4$  & 4.54&1  & 2.09 $\times 10^{13}$, 4.18 $\times 10^{13}$, 6.26 $\times 10^{13}$& 47.51\\
		GJ 3138 b         & M0     & 0.0197 &  0.68  & 0.5  & 0.0056&3.41 $\times 10^4$  & 28.48&1 & 4.03 $\times 10^{14}$, 8.07 $\times 10^{14}$, 1.21 $\times 10^{15}$ & 23.33\\
		GJ 3470 b         & M1.5   & 0.031  &  0.51  & 0.48 & 0.043 &9.31 $\times 10^3$  & 25.2&1  & 3.14 $\times 10^{14}$, 6.29 $\times 10^{14}$, 9.43 $\times 10^{14}$& 23.31\\
		GJ 649 c          & M1.5   & 0.043  &  0.54  & 0.56 & 0.03  &2.30 $\times 10^4$  & 10.34&1 & 2.94 $\times 10^{13}$, 5.87 $\times 10^{13}$, 8.81 $\times 10^{13}$& 12.88\\
		GJ 876 b          & M4     & 0.20832&  0.33  & 0.35 & 2.28  &1.75 $\times 10^4$  & 4.7&$1.90\substack{+0.49 \\ -0.24}$& 8.35 $\times 10^{12}$, 1.06 $\times 10^{13}$, 1.21 $\times 10^{13}$& $11.21\substack{+6.52 \\ -2.64}$\\
		Kepler-42 c       & M5 V   & 0.006  &  0.13  & 0.17 & 0.0011&3.89 $\times 10^4$  & 38.7&1  & 3.49 $\times 10^{14}$, 6.98 $\times 10^{14}$, 1.05 $ \times 10^{15}$& 10.92\\
		GJ 581 e          & M3 V   & 0.02815&  0.31  & 0.29 & 0.0053&1.28 $\times 10^4$  & 6.27&1  & 9.06 $\times 10^{12}$, 1.81 $\times 10^{13}$, 2.72 $\times 10{13}$& 10.81\\
		GJ 1132 b         & M3.5   & 0.0154 &  0.181 & 0.207& 0.0051&2.71 $\times 10^4$  & 12.04&1 & 3.28 $\times 10^{13}$, 6.56 $\times 10^{13}$, 9.84 $\times 10^{13}$& 10.61\\
		GJ 3634 b         & M2.5   & 0.0287 &  0.45  & 0.43 & 0.026 &3.60 $\times 10^4$  & 19.8 &1 & 8.65 $\times 10^{13}$, 1.73 $\times 10^{14}$, 2.59 $\times 10^{14}$& 10.34\\
\enddata
\end{deluxetable*}
\end{longrotatetable}

\section{Discussion and Summary}

Our results have yielded estimates of the steady state radio flux expected from the exoplanets around TRAPPIST-1, Proxima Centauri, and NGTS-1, as well as placing upper limits on the bursts of radio flux which may detected due to extreme conditions in the stellar wind, caused by the factors discussed in Section \ref{trap}. The prediction from our results that exoplanet-induced radio emission should be detectable from M-dwarfs with the VLA, GMRT, and SKA, must take into account whether a signal can be extracted from any background noise in radio observations.  A recent radio survey of M-dwarfs found that for the majority of the sample the flux density was $<$ 100 $\upmu$Jy \citep{mclean2012}. This background emission is approximately one order of magnitude larger than the flux density that our model predicts from exoplanets. However, the known orbital periods of the exoplanets is advantageous when attempting to isolate the signal from the background, as the exoplanet-induced emission will be modulated at the same period, and therefore the light curves from observations can be folded at the orbital period, facilitating detection of the signal. 
 
Although magnetic fields have not been observed at exoplanets, their existence is generally expected \citep{christensen2009}, and a number studies have considered potential methods for detecting their presence, as discussed above. For instance, \citet{scharf2010} proposed that X-ray emissions could probe exoplanets internally through the dynamo-generated magnetic field. Additional, \citet{vidotto2011} studied asymmetries in observations of transiting exoplanets, which may be an indication of bow shocks, which could, in turn, be used to infer the presence of intrinsic magnetic fields. In studies considering ECMI emission directly from the planet, the magnetic field strength may be constrained by the frequency of the radio emission \citep{hess2011a}.  This is not possible for terrestrial planets with weaker magnetic fields, owing to the ionospheric cutoff frequency as discussed earlier, thus hindering the probing of such planetary fields, and strengthening the case for future space-based observatories to access low-frequencies \citep{burkhart2017}. Multi-planet systems, such as TRAPPIST-1 provide an important opportunity to identify the presence of exoplanetary magnetic fields. The poorly constrained nature of a number of parameters, such as the stellar magnetic field strength and mass outflow rate, means that a flux density measurement of a single exoplanet-interaction cannot determine whether that planet possesses an intrinsic magnetic field.  Multiple sources of radio emission, however, from the same system, where factors such as stellar magnetic field, XUV luminosity and stellar wind should vary in a well understood manner, could be compared with the profiles we predict from Figure \ref{fig:trap_derived} to constrain the relative differences in the magnetic field strength of the planets.  This factor distinguishes TRAPPIST-1 from the other systems examined in this study, and for this reason we suggest it as a priority focus for future radio observations.  

Although the flux density predictions of our results for NGTS-1 are similar in magnitude to the predictions for TRAPPIST-1 and Proxima Centauri, we note that NGTS-1 is a much more distant star than the other two case studies (224 pc compared with 12.1 and 1.3 pc of TRAPPIST-1 and Proxima Centauri respectively). Discovery of a similar system to NGTS-1, comprising a hot Jupiter in close orbit around an M-dwarf, but closer to the Solar System, would have a greater probability of radio detection, due to a higher flux density.

It is generally a necessary condition that there is a small separation distance between star and exoplanet (typically $\le 0.2$ AU) for this method to bear results, for a number of reasons.  Most crucially, the sub-Alfvenic condition is typically only satisfied for close orbits.  Secondly, equation (\ref{poyn}) shows that $S_{\mathrm{tot}} \propto v_0^2 \propto v_{\mathrm{orb}}^2$, and hence the total Poynting flux is very sensitive to orbital distance, as illustrated in Table \ref{tab:mdwarfs} where the majority of semi-major axes are $<$ 0.1 AU. Thirdly, as shown in Figure \ref{fig:schematic}, one of the Alfv\'{e}n wings points back towards the star.  However, if the azimuthal component of the stellar wind is sufficiently large, as occurs at greater orbital distance, then both Alfv\'{e}n wings may be produced at angles directed away from the star.  

The results of our study investigating the coherent radio emission generated from sub-Alfv\'{e}nic interaction between exoplanets and M-dwarfs have shown that such emission should be widely detectable currently with the VLA, with the GMRT following its imminent upgrade, and in the future with SKA. This work focused on radio emission originating from near the surface of the star, to overcome the issue of ionospheric blocking of the low frequency signals expected directly from planets, despite the fact that radio flux density of exoplanet origin would be greater than flux from the stellar  surface due to the comparatively weak planetary magnetic fields. In order to access the low frequency radio emission from terrestrial exoplanets, observations must be taken from above the ionosphere.  To this end, \citet{zarka2012} examined the prospects of future radio arrays on the lunar surface.  Alternatively, \citep{rajan2016} has proposed a space-based array of CubeSats for low-frequency radio observations. Future measurements of properties such as the magnetic field strength, rotation period, XUV luminosity and mass outflow rate of M-dwarfs would greatly improve the accuracy of the flux density predictions in this study, and we also suggest that this work is extended in future beyond M-dwarfs to other main sequence stars hosting exoplanets.
 
\acknowledgments

ST was supported by an STFC Quota Studentship.  JDN was supported by STFC grant ST/K001000/1.

%% The reference list follows the main body and any appendices.
%% Use LaTeX's thebibliography environment to mark up your reference list.
%% Note \begin{thebibliography} is followed by an empty set of
%% curly braces.  If you forget this, LaTeX will generate the error
%% "Perhaps a missing \item?".
%%
%% thebibliography produces citations in the text using \bibitem-\cite
%% cross-referencing. Each reference is preceded by a
%% \bibitem command that defines in curly braces the KEY that corresponds
%% to the KEY in the \cite commands (see the first section above).
%% Make sure that you provide a unique KEY for every \bibitem or else the
%% paper will not LaTeX. The square brackets should contain
%% the citation text that LaTeX will insert in
%% place of the \cite commands.

%% We have used macros to produce journal name abbreviations.
%% \aastex provides a number of these for the more frequently-cited journals.
%% See the Author Guide for a list of them.

%% Note that the style of the \bibitem labels (in []) is slightly
%% different from previous examples.  The natbib system solves a host
%% of citation expression problems, but it is necessary to clearly
%% delimit the year from the author name used in the citation.
%% See the natbib documentation for more details and options.

%\begin{thebibliography}{}

\bibstyle{aasjournal}
\bibliography{mybibexo} % if your bibtex file is called example.bib

\begin{thebibliography}{}
\expandafter\ifx\csname natexlab\endcsname\relax\def\natexlab#1{#1}\fi
\providecommand{\url}[1]{\href{#1}{#1}}

\bibitem[{Anglada-Escud{\'e} {et~al.}(2016)Anglada-Escud{\'e}, Amado, Barnes,
  Berdi{\~n}as, Butler, Coleman, de~La~Cueva, Dreizler, Endl, Giesers,
  {et~al.}}]{anglada2016}
Anglada-Escud{\'e}, G., Amado, P.~J., Barnes, J., {et~al.} 2016, Nature, 536,
  437

\bibitem[{Bayliss {et~al.}(2017)Bayliss, Gillen, Eigmuller, McCormac,
  Alexander, Armstron, Booth, Bouchy, \& Burleigh}]{bayliss2017}
Bayliss, D., Gillen, E., Eigmuller, P., {et~al.} 2017, arXiv:1710.11099, in
  pres

\bibitem[{Bell {et~al.}(2016)Bell, Lynch, Kaplan, Murphy, Gaensler,
  Hurley-Walker, Hancock, Lenc, Callingham, Dwarakanath, {et~al.}}]{bell2016}
Bell, M., Lynch, C., Kaplan, D., {et~al.} 2016, The Astronomer's Telegram, 9465

\bibitem[{Bonfond {et~al.}(2009)Bonfond, Grodent, G{\'e}rard, Radioti, Dols,
  Delamere, \& Clarke}]{bonfond2009}
Bonfond, B., Grodent, D., G{\'e}rard, J.-C., {et~al.} 2009, Journal of
  Geophysical Research: Space Physics, 114, A7

\bibitem[{Burkhart \& Loeb(2017)}]{burkhart2017}
Burkhart, B., \& Loeb, A. 2017, arXiv:1706.07038, in pres

\bibitem[{Burrows {et~al.}(1997)Burrows, Marley, Hubbard, Lunine, Guillot,
  Saumon, Freedman, Sudarsky, \& Sharp}]{burrows1997}
Burrows, A., Marley, M., Hubbard, W.~B., {et~al.} 1997, The Astrophysical
  Journal, 491, 856

\bibitem[{Busse(1976)}]{busse1976}
Busse, F. 1976, Physics of the Earth and Planetary Interiors, 12, 350

\bibitem[{Carroll \& Ostlie(2007)}]{carroll2007}
Carroll, B.~W., \& Ostlie, D.~A. 2007, An introduction to modern astrophysics
  (Addison-Wesley, San Francisco)

\bibitem[{Christensen {et~al.}(2009)Christensen, Holzwarth, \&
  Reiners}]{christensen2009}
Christensen, U.~R., Holzwarth, V., \& Reiners, A. 2009, Nature, 457, 167

\bibitem[{Clarke {et~al.}(2004)Clarke, Grodent, Cowley, Bunce, Zarka,
  Connerney, \& Satoh}]{clarke2004}
Clarke, J.~T., Grodent, D., Cowley, S.~W., {et~al.} 2004, Jupiter. The planet,
  satellites and magnetosphere, 1, 639

\bibitem[{Clarke {et~al.}(1996)Clarke, Ballester, Trauger, Evans, Connerney,
  Stapelfeldt, Crisp, Feldman, Burrows, Casertano, {et~al.}}]{clarke1996}
Clarke, J.~T., Ballester, G.~E., Trauger, J., {et~al.} 1996, Science, 274, 404

\bibitem[{Cohen {et~al.}(2015)Cohen, Ma, Drake, Glocer, Garraffo, Bell, \&
  Gombosi}]{cohen2015}
Cohen, O., Ma, Y., Drake, J., {et~al.} 2015, The Astrophysical Journal, 806, 41

\bibitem[{Cranmer(2004)}]{cranmer2004}
Cranmer, S.~R. 2004, American Journal of Physics, 72, 1397

\bibitem[{Curtis \& Ness(1986)}]{curtis1986}
Curtis, S., \& Ness, N. 1986, Journal of Geophysical Research: Space Physics,
  91, 11003

\bibitem[{Ergun {et~al.}(2000)Ergun, Carlson, McFadden, Delory, Strangeway, \&
  Pritchett}]{ergun2000}
Ergun, R., Carlson, C., McFadden, J., {et~al.} 2000, The Astrophysical Journal,
  538, 456

\bibitem[{Farrell {et~al.}(2004)Farrell, Lazio, Zarka, Bastian, Desch, \&
  Ryabov}]{farrell2004}
Farrell, W., Lazio, T., Zarka, P., {et~al.} 2004, Planetary and Space Science,
  52, 1469

\bibitem[{Garraffo {et~al.}(2016)Garraffo, Drake, \& Cohen}]{garraffo2016}
Garraffo, C., Drake, J.~J., \& Cohen, O. 2016, The Astrophysical Journal
  Letters, 833, L4

\bibitem[{Garraffo {et~al.}(2017)Garraffo, Drake, Cohen, Alvarado-Gómez, \&
  Moschou}]{garraffo2017}
Garraffo, C., Drake, J.~J., Cohen, O., Alvarado-Gómez, J.~D., \& Moschou,
  S.~P. 2017, The Astrophysical Journal Letters, 843, L33

\bibitem[{G{\'e}rard {et~al.}(2006)G{\'e}rard, Saglam, Grodent, \&
  Clarke}]{gerard2006}
G{\'e}rard, J.-C., Saglam, A., Grodent, D., \& Clarke, J.~T. 2006, Journal of
  Geophysical Research: Space Physics, 111, A4

\bibitem[{Giampapa {et~al.}(1996)Giampapa, Rosner, Kashyap, Fleming, Schmitt,
  \& Bookbinder}]{giampapa1996}
Giampapa, M.~S., Rosner, R., Kashyap, V., {et~al.} 1996, The Astrophysical
  Journal, 463, 707

\bibitem[{Gillon {et~al.}(2016)Gillon, Jehin, Lederer, Delrez, de~Wit,
  Burdanov, Van~Grootel, Burgasser, Triaud, Opitom, {et~al.}}]{gillon2016}
Gillon, M., Jehin, E., Lederer, S.~M., {et~al.} 2016, Nature, 533, 221

\bibitem[{Gillon {et~al.}(2017)Gillon, Triaud, Demory, Jehin, Agol, Deck,
  Lederer, De~Wit, Burdanov, Ingalls, {et~al.}}]{gillon2017}
Gillon, M., Triaud, A.~H., Demory, B.-O., {et~al.} 2017, Nature, 542, 456

\bibitem[{Grie{\ss}meier {et~al.}(2007)Grie{\ss}meier, Zarka, \&
  Spreeuw}]{griessmeier2007}
Grie{\ss}meier, J.-M., Zarka, P., \& Spreeuw, H. 2007, Astronomy \&
  Astrophysics, 475, 359

\bibitem[{Gupta {et~al.}(2017)Gupta, Ajithkumar, Kale, Nayak, Sabhapathy,
  Sureshkumar, Swami, Chengalur, Ghosh, Ishwara-Chandra, {et~al.}}]{gupta2017}
Gupta, Y., Ajithkumar, B., Kale, H., {et~al.} 2017, Current Science, 113, 707

\bibitem[{Hallinan {et~al.}(2017)Hallinan, Corsi, Mooley, Hotokezaka, Nakar,
  Kasliwal, Kaplan, Frail, Myers, Murphy, {et~al.}}]{hallinan2017}
Hallinan, G., Corsi, A., Mooley, K., {et~al.} 2017, Science, eaap9855

\bibitem[{Hess {et~al.}(2011)Hess, Delamere, Dols, \& Ray}]{hess2011b}
Hess, S., Delamere, P., Dols, V., \& Ray, L. 2011, Journal of Geophysical
  Research: Space Physics, 116

\bibitem[{Hess \& Zarka(2011)}]{hess2011a}
Hess, S., \& Zarka, P. 2011, Astronomy \& Astrophysics, 531, A29

\bibitem[{Imai {et~al.}(2008)Imai, Imai, Higgins, \& Thieman}]{imai2008}
Imai, M., Imai, K., Higgins, C.~A., \& Thieman, J.~R. 2008, Geophysical
  Research Letters, 35, L17103

\bibitem[{Jacobsen {et~al.}(2007)Jacobsen, Neubauer, Saur, \&
  Schilling}]{jacobsen2007}
Jacobsen, S., Neubauer, F., Saur, J., \& Schilling, N. 2007, Geophysical
  research letters, 34, L10202

\bibitem[{Jones \& Su(2008)}]{jones2008}
Jones, S., \& Su, Y.-J. 2008, Journal of Geophysical Research: Space Physics,
  113

\bibitem[{Khodachenko {et~al.}(2007)Khodachenko, Ribas, Lammer, Grie{\ss}meier,
  Leitner, Selsis, Eiroa, Hanslmeier, Biernat, Farrugia,
  {et~al.}}]{khodachenko2007}
Khodachenko, M.~L., Ribas, I., Lammer, H., {et~al.} 2007, Astrobiology, 7, 167

\bibitem[{Kivelson {et~al.}(2004)Kivelson, Bagenal, Kurth, Neubauer, Paranicas,
  \& Saur}]{kivelson2004}
Kivelson, M.~G., Bagenal, F., Kurth, W.~S., {et~al.} 2004, Jupiter: The planet,
  satellites and magnetosphere, 513

\bibitem[{Lammer {et~al.}(2009)Lammer, Bredeh{\"o}ft, Coustenis, Khodachenko,
  Kaltenegger, Grasset, Prieur, Raulin, Ehrenfreund, Yamauchi,
  {et~al.}}]{lammer2009}
Lammer, H., Bredeh{\"o}ft, J., Coustenis, A., {et~al.} 2009, Astronomy and
  Astrophysics Review, 17, 181

\bibitem[{Lamy {et~al.}(2011)Lamy, Cecconi, Zarka, Canu, Schippers, Kurth,
  Mutel, Gurnett, Menietti, \& Louarn}]{lamy2011}
Lamy, L., Cecconi, B., Zarka, P., {et~al.} 2011, Journal of Geophysical
  Research: Space Physics, 116, 2156

\bibitem[{Lazio {et~al.}(2004)Lazio, Farrell, Dietrick, Greenlees, Hogan,
  Jones, \& Hennig}]{lazio2004}
Lazio, T., Farrell, W., Dietrick, J., {et~al.} 2004, The Astrophysical Journal,
  612, 511

\bibitem[{Lazio {et~al.}(2009)Lazio, Carmichael, Clark, Elkins, Gudmundsen,
  Mott, Szwajkowski, \& Hennig}]{lazio2009}
Lazio, T. J.~W., Carmichael, S., Clark, J., {et~al.} 2009, The Astronomical
  Journal, 139, 96

\bibitem[{Luger {et~al.}(2017{\natexlab{a}})Luger, Lustig-Yaeger, Fleming,
  Tilley, Agol, Meadows, Deitrick, \& Barnes}]{luger2017b}
Luger, R., Lustig-Yaeger, J., Fleming, D.~P., {et~al.} 2017{\natexlab{a}}, The
  Astrophysical Journal, 837, 63

\bibitem[{Luger {et~al.}(2017{\natexlab{b}})Luger, Sestovic, Kruse, Grimm,
  Demory, Agol, Bolmont, Fabrycky, Fernandes, Van~Grootel,
  {et~al.}}]{luger2017a}
Luger, R., Sestovic, M., Kruse, E., {et~al.} 2017{\natexlab{b}}, Nature
  Astronomy, 1, 0129

\bibitem[{Lynch {et~al.}(2015)Lynch, Mutel, \& G{\"u}del}]{lynch2015}
Lynch, C., Mutel, R., \& G{\"u}del, M. 2015, The Astrophysical Journal, 802,
  106

\bibitem[{McLean {et~al.}(2012)McLean, Berger, \& Reiners}]{mclean2012}
McLean, M., Berger, E., \& Reiners, A. 2012, The Astrophysical Journal, 746, 23

\bibitem[{McQuillan {et~al.}(2013)McQuillan, Aigrain, \& Mazeh}]{mcquillan2013}
McQuillan, A., Aigrain, S., \& Mazeh, T. 2013, Monthly Notices of the Royal
  Astronomical Society, 432, 1203

\bibitem[{Mead \& Beard(1964)}]{mead1964}
Mead, G.~D., \& Beard, D.~B. 1964, Journal of Geophysical Research, 69, 1169

\bibitem[{Mizutani {et~al.}(1992)Mizutani, Yamamoto, \&
  Fujimura}]{mizutani1992}
Mizutani, H., Yamamoto, T., \& Fujimura, A. 1992, Advances in Space Research,
  12, 265

\bibitem[{Neubauer(1980)}]{neubauer1980}
Neubauer, F. 1980, Journal of Geophysical Research: Space Physics, 85, 1171

\bibitem[{Neubauer(1998)}]{neubauer1998}
Neubauer, F.~M. 1998, Journal of Geophysical Research: Planets, 103, 19843

\bibitem[{Nichols(2012)}]{nichols2012candidates}
Nichols, J. 2012, Monthly Notices of the Royal Astronomical Society: Letters,
  427, L75

\bibitem[{Nichols \& Milan(2016)}]{nichols2016}
Nichols, J., \& Milan, S. 2016, Monthly Notices of the Royal Astronomical
  Society, 461, 2353

\bibitem[{Nichols(2011)}]{nichols2011}
Nichols, J.~D. 2011, Monthly Notices of the Royal Astronomical Society, 414,
  2125

\bibitem[{Parker(1958)}]{parker1958}
Parker, E.~N. 1958, The Astrophysical Journal, 128, 664

\bibitem[{Prang{\'e} {et~al.}(1996)Prang{\'e}, Rego, Southwood, Zarka, Miller,
  \& Ip}]{prange1996}
Prang{\'e}, R., Rego, D., Southwood, D., {et~al.} 1996, Nature, 379, 323

\bibitem[{Rajan {et~al.}(2016)Rajan, Boonstra, Bentum, Klein-Wolt, Belien,
  Arts, Saks, \& van~der Veen}]{rajan2016}
Rajan, R.~T., Boonstra, A.-J., Bentum, M., {et~al.} 2016, Experimental
  Astronomy, 41, 271

\bibitem[{Ray \& Hess(2008)}]{ray2008}
Ray, L., \& Hess, S. 2008, Journal of Geophysical Research: Space Physics, 113

\bibitem[{Reiners \& Basri(2007)}]{reiners2007}
Reiners, A., \& Basri, G. 2007, The Astrophysical Journal, 656, 1121

\bibitem[{Reiners \& Basri(2008)}]{reiners2008}
---. 2008, Astronomy \& Astrophysics, 489, L45

\bibitem[{Reiners \& Basri(2010)}]{reiners2010}
---. 2010, The Astrophysical Journal, 710, 924

\bibitem[{Reiners {et~al.}(2009)Reiners, Basri, \& Christensen}]{reiners2009}
Reiners, A., Basri, G., \& Christensen, U.~R. 2009, The Astrophysical Journal,
  697, 373

\bibitem[{Reiners \& Christensen(2010)}]{reiners&christensen2010}
Reiners, A., \& Christensen, U.~R. 2010, Astronomy \& Astrophysics, 522, A13

\bibitem[{Sano(1993)}]{sano1993}
Sano, Y. 1993, Journal of geomagnetism and geoelectricity, 45, 65

\bibitem[{Saur {et~al.}(2013)Saur, Grambusch, Duling, Neubauer, \&
  Simon}]{saur2013}
Saur, J., Grambusch, T., Duling, S., Neubauer, F., \& Simon, S. 2013, Astronomy
  \& Astrophysics, 552, A119

\bibitem[{Saur {et~al.}(2004)Saur, Neubauer, Connerney, Zarka, \&
  Kivelson}]{saur2004}
Saur, J., Neubauer, F.~M., Connerney, J., Zarka, P., \& Kivelson, M.~G. 2004,
  Jupiter: The Planet, Satellites and Magnetosphere, 1, 537

\bibitem[{Saur {et~al.}(1999)Saur, Neubauer, Strobel, \& Summers}]{saur1999}
Saur, J., Neubauer, F.~M., Strobel, D.~F., \& Summers, M.~E. 1999, Journal of
  Geophysical Research: Space Physics, 104, 25105

\bibitem[{Scharf(2010)}]{scharf2010}
Scharf, C.~A. 2010, The Astrophysical Journal, 722, 1547

\bibitem[{Schmitt {et~al.}(1990)Schmitt, Collura, Sciortino, Vaiana,
  Harnden~Jr, \& Rosner}]{schmitt1990}
Schmitt, J., Collura, A., Sciortino, S., {et~al.} 1990, The Astrophysical
  Journal, 365, 704

\bibitem[{Seager(2013)}]{seager2013}
Seager, S. 2013, Science, 340, 577

\bibitem[{Stevenson(1983)}]{stevenson1983}
Stevenson, D. 1983, Reports on Progress in Physics, 46, 555

\bibitem[{Tingay {et~al.}(2013)Tingay, Goeke, Bowman, Emrich, Ord, Mitchell,
  Morales, Booler, Crosse, Wayth, {et~al.}}]{tingay2013}
Tingay, S., Goeke, R., Bowman, J.~D., {et~al.} 2013, Publications of the
  Astronomical Society of Australia, 30

\bibitem[{Treumann(2006)}]{treumann2006}
Treumann, R.~A. 2006, The Astronomy and Astrophysics Review, 13, 229

\bibitem[{Van~Weeren {et~al.}(2014)Van~Weeren, Williams, Tasse, R{\"o}ttgering,
  Rafferty, Van Der~Tol, Heald, White, Shulevski, Best, {et~al.}}]{van2014}
Van~Weeren, R.~J., Williams, W.~L., Tasse, C., {et~al.} 2014, The Astrophysical
  Journal, 793, 82

\bibitem[{Vidotto {et~al.}(2011)Vidotto, Jardine, \& Helling}]{vidotto2011}
Vidotto, A., Jardine, M., \& Helling, C. 2011, Monthly Notices of the Royal
  Astronomical Society: Letters, 411, L46

\bibitem[{Vidotto {et~al.}(2013)Vidotto, Jardine, Morin, Donati, Lang, \&
  Russell}]{vidotto2013}
Vidotto, A., Jardine, M., Morin, J., {et~al.} 2013, Astronomy \& Astrophysics,
  557, A67

\bibitem[{Wang {et~al.}(2017)Wang, Wu, Barclay, \& Laughlin}]{wang2017}
Wang, S., Wu, D.-H., Barclay, T., \& Laughlin, G.~P. 2017, arXiv preprint
  arXiv:1704.04290

\bibitem[{Wannawichian {et~al.}(2010)Wannawichian, Clarke, \&
  Nichols}]{wannawichian2010}
Wannawichian, S., Clarke, J., \& Nichols, J. 2010, Journal of Geophysical
  Research: Space Physics, 115

\bibitem[{Wheatley {et~al.}(2017)Wheatley, Louden, Bourrier, Ehrenreich, \&
  Gillon}]{wheatley2017}
Wheatley, P.~J., Louden, T., Bourrier, V., Ehrenreich, D., \& Gillon, M. 2017,
  Monthly Notices of the Royal Astronomical Society: Letters, 465, L74

\bibitem[{Wright \& Schwartz(1989)}]{wright1989}
Wright, A.~N., \& Schwartz, S.~J. 1989, Journal of Geophysical Research: Space
  Physics, 94, 3749

\bibitem[{Wu \& Lee(1979)}]{wu1979}
Wu, C., \& Lee, L. 1979, The Astrophysical Journal, 230, 621

\bibitem[{Zarka(1992)}]{zarka1992}
Zarka, P. 1992, Advances in Space Research, 12, 99

\bibitem[{Zarka(1998)}]{zarka1998}
---. 1998, Journal of Geophysical Research: Planets, 103, 20159

\bibitem[{Zarka(2007)}]{zarka2007}
---. 2007, Planetary and Space Science, 55, 598

\bibitem[{Zarka {et~al.}(2004)Zarka, Cecconi, \& Kurth}]{zarka2004}
Zarka, P., Cecconi, B., \& Kurth, W.~S. 2004, Journal of Geophysical Research:
  Space Physics, 109, A9

\bibitem[{Zarka {et~al.}(2015)Zarka, Lazio, \& Hallinan}]{zarka2015}
Zarka, P., Lazio, J., \& Hallinan, G. 2015, Advancing Astrophysics with the
  Square Kilometre Array (AASKA14)

\bibitem[{Zarka {et~al.}(2012)Zarka, Bougeret, Briand, Cecconi, Falcke, Girard,
  Grie{\ss}meier, Hess, Klein-Wolt, Konovalenko, {et~al.}}]{zarka2012}
Zarka, P., Bougeret, J.-L., Briand, C., {et~al.} 2012, Planetary and Space
  Science, 74, 156

\end{thebibliography}

%\end{thebibliography}

%% This command is needed to show the entire author+affilation list when
%% the collaboration and author truncation commands are used.  It has to
%% go at the end of the manuscript.
%\allauthors

%% Include this line if you are using the \added, \replaced, \deleted
%% commands to see a summary list of all changes at the end of the article.
%\listofchanges

\appendix
\section{Calculating interaction strength factor $\bar{\upalpha}$}  \label{alpha}

The plasma flow-obstacle interaction strength factor $\bar{\upalpha}$ can be determined by considering the ionospheric Pedersen conductance in the case of either a magnetised or unmagnetised planet. \citet{neubauer1998} and \citet{saur1999} showed that $\bar{\upalpha}$ can be approximated by

\begin{equation} \label{alpha}
\bar{\upalpha} = \frac{\Sigma_{\mathrm{P}}}{\Sigma_{\mathrm{P}} + 2\Sigma_{\mathrm{A}}},
\end{equation}
where $\Sigma_{\mathrm{P}}$ is the ionospheric Pedersen conductance, and $\Sigma_{\mathrm{A}}$ is the Alfv\'{e}n conductance, given by 

\begin{equation} \label{alf_cond}
\Sigma_{\mathrm{A}} = \frac{1}{\upmu_0 v_{\mathrm{A}}},
\end{equation}
where the Alfv\'{e}n speed $v_{\mathrm {A}}$ in a stellar wind of mass density $\rho_{\mathrm{sw}}$ is 

\begin{equation} \label{alfven}
v_{\mathrm{A}} = \frac{B_{\mathrm{sw}}}{\left(\upmu_0 \rho_{\mathrm{sw}}\right)^{1/2}}.
\end{equation}
Here, the ionospheric Pedersen conductivity is estimated using the empirical power law  of \citet{nichols2016}, i.e.

\begin{equation} \label{ped}
\Sigma_{\mathrm{P}} = \upkappa \left( \frac{d}{1 \mathrm{AU}} \right)^{\uplambda} \left( \frac{B_J}{B_p} \right) \left( \frac{L_{\mathrm{XUV}}}{L_{\mathrm{XUV \odot}}} \right)^{\upmu} \mathrm{mho},
\end{equation}
where $d$ is the orbital distance of the planet, $B_p$ is the equatorial exoplanetary magnetic field strength, $B_J$ is the surface field strength at Jupiter, $L_{\mathrm{XUV}}$ is the stellar XUV luminosity, and $L_{\mathrm{XUV \odot}}$ is the solar value. The constants in equation (\ref{ped}) take the values $\upkappa$ = 15.475, $\uplambda$ = -2.082, and $\upmu$ = 0.5. 

For M-dwarfs we assume a value of $L_{\mathrm{XUV}} = L_{\mathrm{XUV \odot}}$, consistent with X-ray observations of TRAPPIST-1 \citep{wheatley2017}.  In all the cases we consider of close-orbiting exoplanets, $\Sigma_{\mathrm{P}} \gg \Sigma_{\mathrm{A}}$, therefore $\bar{\upalpha} \simeq 1$.  We note that for an unmagnetised or weakly magnetised planet this approximation is also valid, since in this case $\Sigma_{\mathrm{P}}$ approaches infinity.

\section{Model of the stellar wind and magnetic field}

We use an isothermal stellar wind \citep{parker1958} fully parameterised by the sound speed $c_s$ for which we assume a value of 170 km s$^{-1}$, corresponding to a cornal temperature of $\sim$ 2 $\times$ 10$^6$ K, consistent with the temperature adopted by \citet{vidotto2013} in their study of M-dwarf stellar winds. Specifically, we employ \citeauthor{cranmer2004}'s (\citeyear{cranmer2004}) closed-form analytic solution of the isothermal wind equation, given by

\begin{equation} \label{sw}
v_{\mathrm{sw}}^2 = \begin{cases}
- v_c^2W_0[-D(d)] & \text{if} d \le d_c, \\
- v_c^2W_{-1}[-D(d)] & \text{if} d \ge d_c, 
\end{cases}
\end{equation} 
where $W_0$ and $W_{-1}$ are branches of the Lambert $W$ function, and $D(d)$ is given by 

\begin{equation}
D(d) = \left(\frac{d}{d_c}\right)^{-4} \exp \left[4 \left( 1 - \frac{d_c}{d} \right) -1 \right],
\end{equation}
where $d_c$ is the critical distance at which the stellar wind speed $v_{\mathrm{sw}}$ passes through the sound speed $c_s$, given by

\begin{equation}
d_c = \frac{GM_{\star}}{2 c_s^2}
\end{equation}
for a star of mass $M_{\star}$.

The magnetic field components of the Parker Spiral are given by

\begin{equation} \label{radial}
B_{\mathrm{r}} = B_{\star} \left( \frac{d_0}{d} \right)^2,
\end{equation}
and

\begin{equation} \label{azimuthal}
B_{\phi} = B_{\mathrm{r}} \frac{\Omega_{\star} d}{v_{\mathrm{sw}}},
\end{equation}
where $\Omega_{\star}$ is the stellar rotational velocity, and here we take $d_0 = R_{\star}$, i.e. we assume that the field is radial at the stellar surface.  The resultant IMF magnitude is given by

\begin{equation} \label{mag}
B_{\mathrm{sw}} = \sqrt{\left(B_r^2 + B_{\phi}^2 \right)}.
\end{equation}
The angle $\theta$ between the stellar wind magnetic field and the impinging plasma velocity can hence be defined by 

\begin{equation}
\theta = \arctan \left(\frac{B_{\phi}}{B_r} \right) - \arctan \left(\frac{v_{\mathrm{orb}}}{v_{\mathrm{sw}}} \right).
\end{equation}
Interaction of the stellar wind with an intrinsic planetary magnetic field governs the location of the substellar magnetopause standoff distance $R_{\mathrm{mp}}$, which can be calculated via a consideration of pressure balance, and is given by

\begin{equation} \label{rmp}
\left(\frac{R_{\mathrm{mp}}}{R_p}\right) =  \left( \frac{  k_m^2 B^2_{\mathrm{p}} }{2 \upmu_0 ( p_{\mathrm{dyn \, sw}}  + p_{\mathrm{th \, sw}} +  B^2_{\mathrm{sw}} / 2 \upmu_0  )         } \right)^{1/6},
\end{equation}
where $k_m$ = 2.44 represents the factor by which the magnetopause currents enhance the magnetospheric magnetic field at the magnetopause for a realistic boundary shape \citep{mead1964}, $p_{\mathrm{th \, sw}}$ is the thermal pressure of the stellar wind, and $p_{\mathrm{dyn \, sw}}$ is the stellar wind dynamic pressure given by

\begin{equation} \label{press}
p_{\mathrm{dyn \, sw}}  = \rho_{\mathrm{sw}} v_0^2,
\end{equation}
where $\rho_{\mathrm{sw}}$ is the density of the stellar wind, which, for a stellar mass loss rate of $\dot{M_{\star}}$ is determined by

\begin{equation} \label{rho}
\rho_{\mathrm{sw}} = \frac{\dot{M_{\star}}}{4 \uppi d^2 v_{\mathrm{sw}}},
\end{equation}
and the corresponding plasma number density is given by

\begin{equation} \label{number}
n_{\mathrm{sw}} = \rho_{\mathrm{sw}} / m_{\mathrm{av}},
\end{equation} 
where we take a Sun-like value of the average particle mass in the stellar wind $m_{\mathrm{av}}$ of 1.92 $\times$ 10$^{-27}$ kg.

\begin{figure*}
	\includegraphics[width=0.8\textwidth]{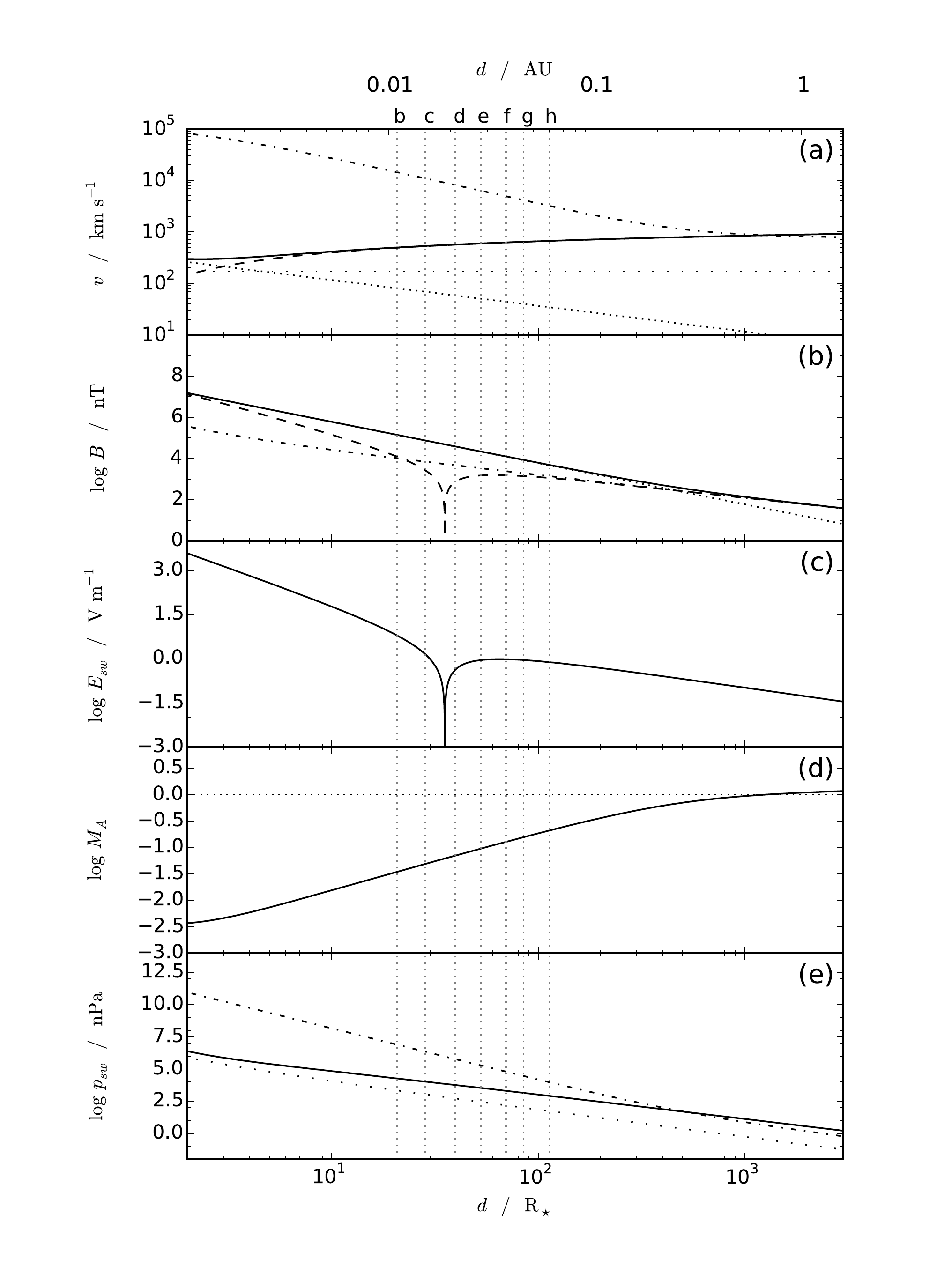}
    \caption{Stellar wind parameters for TRAPPIST-1 versus orbital distance $d$ in units of stellar radii $R_{\star}$ and AU. Panel (a) shows the stellar wind speed $v_{\mathrm{sw}}$ (dashed line), the Keplerian speed (close-dotted line), and the resultant incident stellar wind speed $v_0$ (solid line).  Also shown are the constant sound speed $c_{\mathrm{s}}$ (loose-dotted line), and the Alfv\'{e}n speed $v_{\mathrm{A}}$ (dot-dashed line). Panel (b) shows the stellar magnetic field radial component $B_{\mathrm{r}}$ (dotted line), azimuthal component $B_{\phi}$(dot-dashed line), resultant magnetic field strength $B$ (solid line), and the component perpendicular to the impinging stellar wind $B_{\perp}$ (dashed line).  Panel (c) shows the motional electric field or $E_{\mathrm{sw}}$.  Panel (d) shows the Alfv\'{e}n Mach number $M_{\mathrm{A}}$.  Panel (e) shows the stellar magnetic pressure (dot-dashed line), the dynamic pressure (solid line) and the thermal pressure (loose-dotted line). The vertical dotted lines running through all panels denote the orbital distances of the planets TRAPPIST-1(b) - (h).}
    \label{fig:trap_sw}
\end{figure*}

\begin{figure*}
	\includegraphics[width=0.8\textwidth]{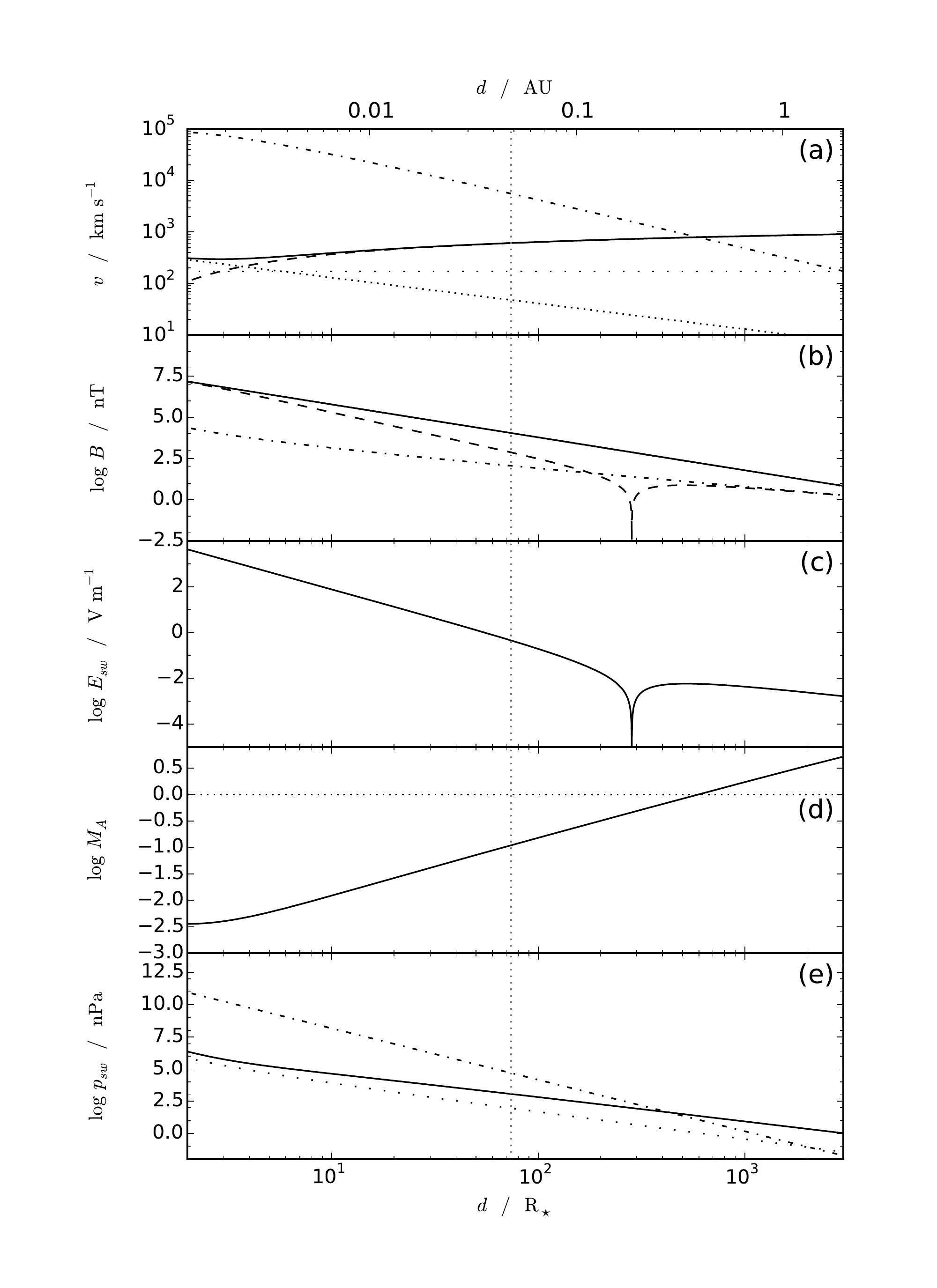}
	
    \caption{As for Figure \ref{fig:trap_sw} but for Proxima Centauri}
    \label{fig:proxima_sw}
\end{figure*}

\begin{figure*}
	\includegraphics[width=0.8\textwidth]{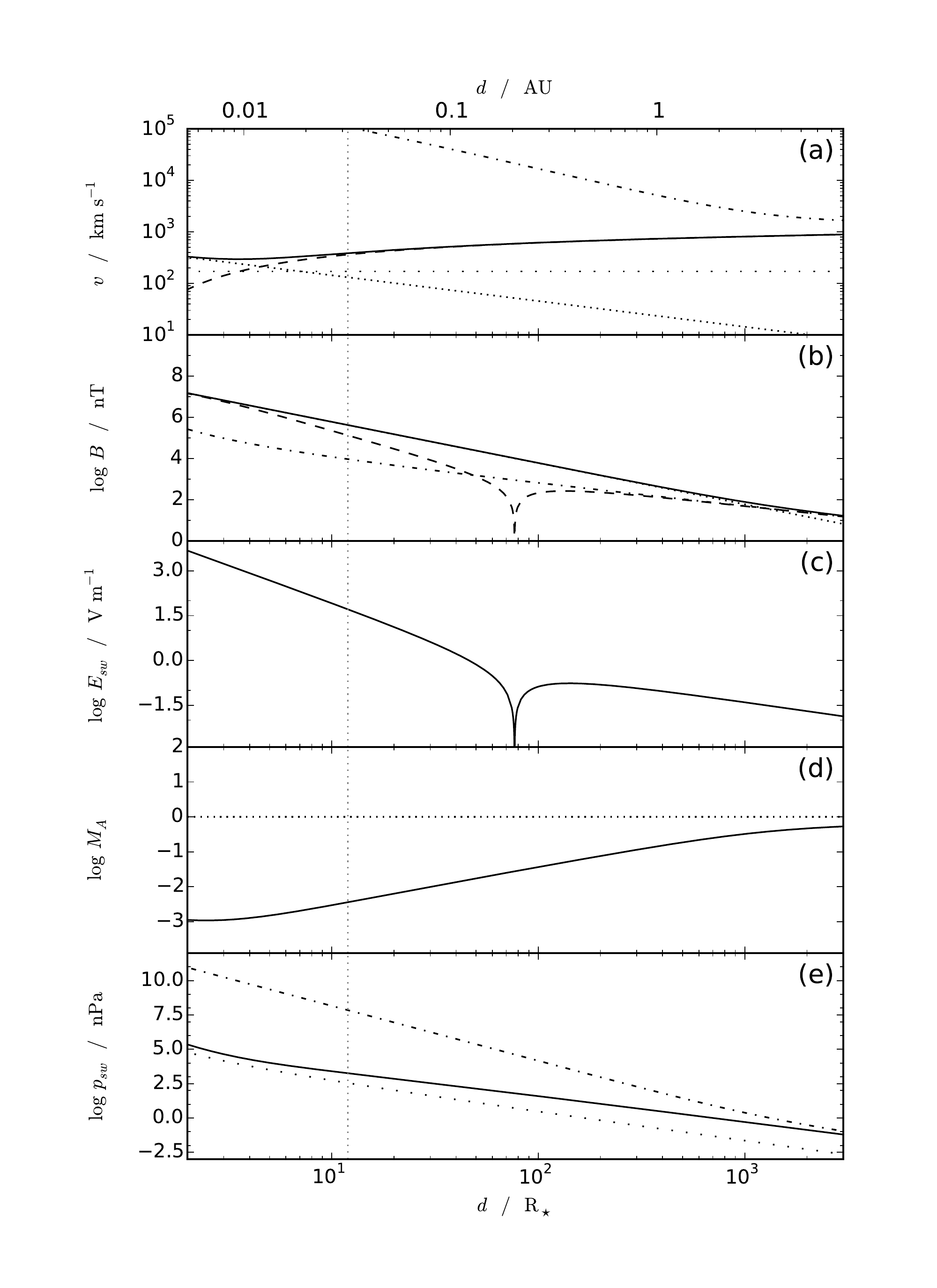}
    \caption{As for Figures \ref{fig:trap_sw} and \ref{fig:proxima_sw} but for NGTS-1.}
    \label{fig:nltt_sw}
\end{figure*}

Figures \ref{fig:trap_sw}, \ref{fig:proxima_sw}, and \ref{fig:nltt_sw} show profiles of various stellar wind parameters for our case studies of TRAPPIST-1, Proxima Centauri, and NGTS-1 respectively.

\end{document}